\documentclass[useAMS,usenatbib]{mn2e}
\usepackage{graphicx,amsmath,natbib}
\usepackage{subfigure}

\title[Collimation and rotation of disk winds]{Controlling the collimation
and rotation of hydromagnetic disk winds}

\author[Pudritz, R.E., Rogers, C.S., \& Ouyed, R.]{Ralph E. Pudritz$^{1,2}$, 
Conrad S. Rogers$^1$, and 
Rachid Ouyed$^3$ \\ $^1$ Department of Physics and Astronomy, McMaster University,
Hamilton, ON L8S 4M1, Canada \\ $^2$ Origins Institute, McMaster University,
Hamilton, ON L8S 4M1, Canada \\ $^3$ Department of Physics and Astronomy, University
of Calgary, Calgary, AB T2N 1N4, Canada} 

\date{Released 2004 Xxxxx XX}

\pagerange{\pageref{firstpage}--\pageref{lastpage}} \pubyear{2004}

\def\LaTeX{L\kern-.36em\raise.3ex\hbox{a}\kern-.15em
    T\kern-.1667em\lower.7ex\hbox{E}\kern-.125emX}

\begin{document}

\label{firstpage}

\maketitle
\begin{abstract}

We present a comprehensive set of axisymmetric, time-dependent simulations
of jets from Keplerian disks whose mass loading 
as a function of 
disk radius is systematically changed.  For a reasonable model
for the density structure and injection speed of the underlying
accretion disk, mass loading is determined by the radial structure of 
the disk's magnetic field structure.  We vary this structure 
by using four different magnetic field configurations, ranging
from the "potential" configuration (Ouyed \& Pudritz 1997), to 
the increasingly more steeply falling Blandford \& Payne (1982) and 
Pelletier \& Pudritz (1992) models, and ending with a quite
steeply raked configuration that bears similarities to the 
Shu X-wind model.  We find that the 
radial distribution of the mass load has 
a profound effect on both the rotational profile of the underlying
jet as well as the degree of collimation of its outflow velocity and 
magnetic field lines.  These four models have systematic differences
in the power-law rotation profiles of jet material far from 
the source; $v_{\phi(r)} 
\propto r^{a} $ ranging over $-0.46 \ge a \ge -0.76$.  
We show analytically, and confirm by our simulations, that 
the collimation of a jet depends on its radial current
distribution, which in turn is prescribed by the mass load.
Models with steeply descending mass loads have strong toroidal
fields, and these collimate to cylinders (this includes the Ouyed-Pudritz
 and Blandford-Payne outflows).  On the other hand, the more gradually descending
mass load profiles (the PP92 and monopolar distributions) have
weaker toroidal fields, and these result in wide-angle outflows
with parabolic collimation.  We also present detailed structural
information about jets such as their radial profiles of jet density,
toroidal magnetic field, and poloidal jet speed, as well as an
analysis of the bulk energetics of our different simulations. 
Our results are in excellent agreement with the predictions of 
asymptotic collimation for axisymmetric, stationary jets. 

\end{abstract}

\section{Introduction} 

Hydromagnetic disk winds
provide what is arguably the most comprehensive 
quantitative picture for the origin
of astrophysical jets in protostellar systems.
An extensive body of theoretical work as well as 
a wide range of numerical simulations, using very 
different codes, robustly demonstrate
that a centrifugally driven wind from an
accretion disk can efficiently extract 
its angular momentum and gravitational potential energy. 
Disk winds provide a potentially universal
mechanism for the origins of jets that are observed in both low and high mass 
protostars (eg. reviews; \citet{Richer00,Reipurth01,Zhang05,Shepherd05}) 
as well many other systems such as microquasars (eg. \citet{Mirabel98})
and quasars (e.g. \citet{Begelman84}) because jet properties - such as  
their terminal velocities and wind mass-loss rates -  
naturally scale 
with the depth of the gravitational
potential well of the central object and the accretion
rate into it (eg. reviews; \citet{Spruit96,Konigl00,Pudritz03,Pudritz05}).       

The theory has recently received considerable empirical support 
by HST observations of the rotation of 
several jets associated with 
T-Tauri stars (TTS)  (eg. \citet{Coffey04,Bacciotti00,Bacciotti02}). 
High spectral and spatial resolution observations
have directly measured the rotation of these jets.  
The results suggest that the observed jets might have arised
 from a region of roughly 1 AU in size in the disk.
Moreover, the  angular momentum that jets are inferred to carry 
from this measurement,  
is a significant fraction that must be carried
off to drive the observed accretion rates in the
underlying disks. 
These results seem to agree   with the theoretical predictions of disk
 wind theory (eg., \citet{Anderson03}).  

The observations of jets 
provoke several important questions about
the role of disk winds.
How do the underlying accretion disks control
jet collimation, even to very large scales?  And what determines how rapidly
the observed jets spin and the amount of angular momentum
that they can extract from the disk?   We show in this paper that
both of these important properties of jets are aspects
of the toroidal dynamics of the jet and are controlled
by the same physical process, namely, by the mass  
loading of the disk wind field lines that occurs near 
the surface of the disk.

Consider first the issue of jet collimation. 
The basic mechanism 
has been understood, at least
in principle, for some time.  It involves the 
action of a hoop stress 
that arises from the generation of a toroidal magnetic field  
in the rotating, magnetized flow,  
which pinches the flow
towards the outflow axis. 
The detailed analysis for the collimation of outflows for 
steady-state jets is extremely difficult since it
involves the solution of the highly nonlinear, Grad-Shafranov
equation.  Progress has been limited to studies of the
asymptotic properties of steady state jets 
\citep{Heyvaerts89} (henceforth HN89), 
or special self-similar models (eg. \citet{Ostriker98,Li95,Trussoni97,Casse00}). 
The general collimation properties of time-dependent jets 
have, so far, defied general
theoretical analysis.

Nature probably produces jets with a range of 
intrinsic collimations, as suggested by the  
observations of molecular outflows. 
Models for observed outflows fall into two general categories: the
jet-driven bow shock picture, and a wind-driven 
shell picture in which the molecular gas is driven by 
an underlying wide-angle wind component
(eg. review \citet{Cabrit97}). 
A survey of molecular outflows by \citet{Lee00} found 
that both mechanisms are needed in order to explain the full 
set of systems observed.

Can the mechanism of jet collimation readily accommodate
both possibilities?
The simulations of disk winds explore a range of initial magnetic
configurations  that either originate on the 
surface of a star and connect with the 
disk (eg. \citet{Hayashi96,Goodson97,Fendt99,Keppens00}),
 within the disk (eg., \citet{Rekowski04}),
or that thread only the disk itself
(eg. \citet{OPS97,Tomisaka98,Kudoh02,Fendt02}).  
In the former case, twisting of the magnetic    
field that initially threads the 
disk beyond the co-rotation radius inflates and then 
disconnects the field from the star 
to produce a steeply decreasing, disk-field. 
In the latter case of purely disk-threaded field, simulations
have a variety of initial magnetic geometries: from  
the split monopole of \citet{Romanova97},  
similar to the highly concentrated
magnetic field of the \citet{Shu00} X-wind configuration,  
to the far more gradual declining or constant magnetic fields 
featured in the potential solutions 
investigated by OPS97.  The familiar, self-similar
disk wind magnetic geometry of Blandford
\& Payne (1982; BP82) is intermediate
between these two extreme regimes with an initial 
poloidal magnetic field on the surface of the disk that falls
off  
with disk radius as $B_{BP82}(r_o) \propto r_o^{-5/4}$.  

All of these simulations show that outflows are
centrifugally driven but that the degree to which they
collimate varies.
Although the
steeply raked, monopolar-like,
magnetic field distributions 
fail to collimate to the outflow axis, they are 
still candidates for jet models.
This is because
the densest part of the outflow is directed along those more polar
lines giving the appearance of "collimated jets" 
(even winds from purely dipole stellar field
produce this structure; \cite{Sakurai85,Matt04a,Matt04b}).  

We show in this paper 
that it is the mass loading of a jet 
- defined as the mass outflow 
rate per unit magnetic flux - 
that determines how well jets are collimated.  
In earlier theory and numerical papers 
we found that low mass loads for jets lead to rapid,
episodic behaviour while more heavily mass-loaded systems tend to 
achieve stationary outflow configurations
(OPS97, Ouyed \& Pudritz 1997b (OP97b), Ouyed \& Pudritz 1999 (OP99)).  
This has also been examined in 
recent work by \citet{Anderson05}.

Regarding the second major issue - angular momentum transport
in the jet - we show that this is related to angular momentum
extraction by the disk wind torque upon the disk.  In steady 
state, the wind torque depends on the strength of the toroidal
magnetic field near the disk surface and this too, we show,
depends on how the wind is mass loaded at the disk surface.

We present a series of numerical simulations of jets
using a much broader range of initial magnetic configurations than 
employed in our earlier papers.
We vary the mass per unit magnetic flux that threads the disk
(the mass loading function) by employing a wide 
range of initial magnetic field distributions across  
our disk model.  Our initial magnetic configurations range from
rather well collimated initial fields (such as the "potential configuration"
of OP97a) and the BP82 configuration, to 
the initially less well collimated configuration of Pelletier-Pudritz (1992; PP92).
The final configuration in our simulation suite features
an even more steeply declining threading field 
with disk radius, and is even more open.  

One of our major findings is that there are 
indeed two different kinds of collimation that
can be achieved by disk winds with different mass loading profiles:
cylindrical collimation and a wide-angle outflow. 
We provide detailed
numerical data that show the radial structure and collimation of
a variety of physical variables, including all components of the 
velocity and magnetic field of the jet, as well as its density
profile.  We show that this agrees with the predictions that 
jet collimation is related to the 
radial structure of the current in these jets, as was
first demonstrated analytically by Heyvaerts \&
 Norman (1989, HN89).  In fact, the wide-angle flows we find correspond
quite well to the case of parabolic collimation predicted in 
this theory.

We also show that the mass loading of jets affects their 
radial angular rotation profiles.  In particular, we find that  
the radial distribution of jet rotation is robust and follows 
radial power-law
relations, $ v_{\phi} \propto r^{a} $, whose index a varies little
from model to model: typically in the range $- 0.46 \ge a \ge - 0.76$ 
depending on the mass-load profile.  
These radial rotation profiles can in principle, be used to 
discriminate between different disk wind models because the
distribution of angular momentum in the wind reflects 
on how it is
mass loaded by the accretion disk at its base. 

We begin this paper by first briefly summarizing some recent 
important observational 
results on the relation between accretion disks and jets from
YSOs (\S 2) and following this up with a brief review of the basic aspects
of hydromagnetic disk wind theory needed to interpret our
numerical results (\S 3).  We then outline our generalized disk wind models
which include simulations of BP82 and PP92,
and other mass loading profiles, in \S4.  We analyze our results in \S 5, and conclude in \S 6.

\section{Observations of jet structure and dynamics}

Of the many important contributions on the issue of jet structure
and dynamics, we 
focus on just a few recent advances that 
theories and 
simulations must address. 
As an example, recent 
detailed high resolution imaging and spectroscopic studies
of jets have revealed their internal dynamical structure in 
sufficient detail to provide useful constraints on 
jet theories and numerical simulations.     
High resolution images of $H_{\alpha}$ and
[OI] emission in the
DG Tau jet taken in various
velocity ranges
from $50 - 450 $ km
s$^{-1}$ in $125$ km s$^{-1}$ wide velocity
bins \citep{Bacciotti02} reveal an "onion-like" velocity structure
wherein 
there is a continuous variation in the velocity of the
jet, the very highest being 
the densest and nearest the jet axis, and the lowest farther away.
The lower speed flow is farther from the axis
and appears to broaden out and become less collimated
with increasing radius in the jet.
The jet can be traced to
within 15 AU (0.1") of the central source.
While earlier papers suggested that jets may have two distinct velocity
components (fast and highly collimated, versus slower and wide-angled components
\citet{Kwan88}), the recent data suggest a continuously varying velocity
structure.

A second important observed property of jets 
is that they rotate.  For DG Tau as an example,  
the flow appears to be
rotating clockwise when viewed from the flow
towards the source \citep{Bacciotti03a,Bacciotti03b}.
The rotation speed is 6 - 15 km s$^{-1}$, depending on the position
in the flow within 110 AU from the source.  These authors
find agreement between these measured rotation speeds with those expected
in a centrifugally driven hydromagnetic wind.  Using the
scalings for disk wind speeds and angular momentum
transport rates outlined in \S 3.1 and \S 3.2,
this outflow is estimated to be
carrying up to 60 \% of the angular momentum loss from
the disk at its footpoint that is needed to drive 
the observed accretion rate.

Finally, as has been noted many times in 
the literature,  
there is a strong correlation between the observed 
mass accretion rate through protostellar disks 
and the mass loss rates carried by their jets.  
Optical jets associated with TTS have typical mass loss
rates that are consistently about a tenth of the
underlying disk accretion rate.
As an example, \citet{Hartmann95} found that for a sample
of 31 TTS, jet mass loss rates 
calculated from the forbidden lines 
in the range $3 \times 10^{-7} < \dot M_j 
< 10^{-10} M_{\odot}$ yr$^{-1}$
with a median value of $10^{-9} M_{\odot} $ yr$^{-1}$ 
(eg. review Calvet 2003). These mass loss rates
are a tenth of the underlying disk accretion rate in the system; 
\begin{equation} \dot M_j / \dot M_a
\simeq 0.1.  
\end{equation}
Theory predicts that 
this is a reflection of the efficient angular momentum extraction
from accretion disks by the wind torque.  

\section{Disk wind theory: jet structure and the role of mass loading}

The theory of hydromagnetic winds originated with  
the early work on winds from rotating magnetized stars
(eg. \citet{Mestel68,Weber67}).   These papers 
developed 1-D, axisymmetric models of hydromagnetic 
flows from rotating stars.  The application of 
this idea to self-similar accretion disks was first carried out 
in the seminal paper by BP82.

Consider the simplest possible description of magnetized, 
rotating, conducting gas of
density $\rho$ that is threaded by a large-scale field.
There are 4 equations of stationary, ideal MHD that 
describe such a system, namely: the conservation  
of mass (continuity equation); 
the equation of motion for gas   undergoing
a pressure gradient force (from the pressure $p$),  
as well as a Lorentz force
(from the field ${\bf B}$) - all within 
in the  gravitational field of a  central
object (whose gravitational potential is $\phi$); 
the induction equation   
for the evolution of the magnetic field in the moving
gas; and finally, the conservation of magnetic flux. 
These equations were written down by Chandrasekhar, Mestel, 
and many others: 

\begin{equation}
 \nabla . ( \rho {\bf v}) = 0 
\end{equation}

\begin{equation}
 \rho {\bf v . \nabla v} = - \nabla p - \rho \nabla \phi
           + { 1 \over 4 \pi} ( {\bf \nabla \times B}) \times {\bf B}
\end{equation}

\begin{equation}
 {\bf \nabla \times (v \times B) } = 0 
\end{equation}

\begin{equation}
 {\bf \nabla . B} = 0 
\end{equation}

\noindent
It is convenient to decompose all magnetic fields and velocity fields
into poloidal and toroidal field components $  {\bf B =  B_p} +  B_{\phi}
{\bf \hat e_{\phi}} $ and $ { \bf v = v_p} + v_{\phi} {\bf \hat e_{\phi}}$.       

The most basic conservation laws that these equations
express are the conservation of mass and magnetic flux 
(equations 2 and 5).  
By comparing the form of these two equations, we see
that the poloidal field and the poloidal mass flux along
the field line must be proportional 
to one another along a field line.  Defining 
a function  
$ k$ that is a constant along a magnetic field line and
which we will call the ``mass load'' of the wind, then;   
\begin{equation}
\rho {\bf v_p} = k {\bf B_p}
\end{equation}
\noindent

This mass load function can be cast in a much more revealing way by noting
that the wind mass loss rate passing through an 
annular section of the flow 
of area $dA$ through the flow is $d \dot M_w = \rho v_p dA$,
while the amount of poloidal magnetic flux through this
same annulus is $d \Phi = B_p dA $.
Thus, the mass load per unit time, per unit magnetic flux
of the wind that is preserved along each
streamline along the flow emanating from the rotor (a disk
in this case) is 
\begin{equation}
k = { \rho v_p \over B_p} = {d \dot M_w \over d \Phi}
\end{equation}

The ultimate source for the
mass load is, of course, the accretion disk.
While there may be several complicated processes in 
the disk that prescribe the function $k$, the stationary
theory summarized in equation (7) states that this mass load
all comes down to the playoff between the gas density, the injection
speed, and the magnetic field strength across the surface of the 
accretion disk.

The strength of the toroidal magnetic field in the 
jet plays an important role
in jet collimation. 
We calculate the strength of this field component
at any point in a stationary,
axisymmetric jet 
by using the poloidal component of  the induction equation
which can be written as,
\begin{equation}
B_{\phi} = {\rho r \over k} ( \Omega - \Omega_o).  
\end{equation}
This result shows that the strength of the toroidal field 
at any point in the flow depends on three things: 
the relative shear of the flow between the point 
in question and the footpoint of the 
field line; the density of the gas (the denser it is the greater
the inertia and hence the greater the toroidal field); and finally
the mass load.    
The larger the value of $k$ along a field line, the smaller
is the value of the toroidal field, and hence the more diminished
role that it plays in collimating the outflow.  
All else being equal, we expect to have stronger toroidal fields
in the portions of an outflow that are fed with 
smaller mass loads.  From this it is clear that
the radial distribution of the mass load $k$ in the 
jet will play a significant role in controlling its collimation. 

Now let us turn our attention 
to the radial rotational profile of 
a hydromagnetic jet.  We derive this from the conservation
of angular momentum equation for axisymmetric, stationary 
flows.   This is described by the  $ \phi $ component
of equation (3), which yields;  

\begin{equation}
 \rho {\bf v_p} . \nabla (r v_{\phi}) = { {\bf B_p} \over 4 \pi} .
                   \nabla (r B_{\phi})  
\end{equation}
Let us apply this equation to two different situations -
(i) the transport of angular momentum along a field line and
(ii) the transport of angular momentum out of a disk, 
by an outflow (which is analyzed in the following subsection). 

By applying  
equation (6) to equation (9) the 
angular momentum equation  
reduces to 
\begin{equation}
 {\bf B_p . \nabla}  (r v_{\phi} - {r B_{\phi} \over 4 \pi k} ) = 0
\end{equation}
\noindent
This equation says that there is a 
conserved quantity,  
\begin{equation}
l = r v_{\phi} - {r B_{\phi} \over 4 \pi k}, 
\end{equation}
along any given field-line
in the flow.  This is 
the total angular momentum per unit mass.   
The form for $l$ reveals that the total angular momentum 
is carried by both the rotating gas (first term) as well
by the twisted field (second term). 
Once again, we see that the mass load plays a significant
role - here by controlling the relative amount of angular momentum that 
is carried by the rotation of material in the jet as compared to 
that carried by the jet's twisted magnetic field.

The conservation of angular momentum given above, can be 
rearranged to find the 
rotational profile of the jet; 
\begin{equation}
r v_{\phi} = { lm^2 -  r^2 \Omega_o \over m^2 - 1}
\end{equation}
where the so-called Alfv\'en Mach number of the flow
is $m = v_p / v_A$ and $v_A = B_p / \sqrt ( 4 \pi \rho)$
is the Alfv\'en speed of the flow.  The Alfv\'en surface
is that point $r = r_A$ at each field line 
in the wind where $m=1$.  The flow may
be regarded as kept in co-rotation with the disk until
this point is reached.  

The actual value that $l$ takes along any field
line is determined by the Alfv\'en surface
where $m=1$;   
\begin{equation}
l(a) = \Omega_o r_A^2 
\end{equation}
For a field line starting at a point $r_o$ on the 
rotor (disk in our case), the Alfv\'en radius is
$r_A(r_o)$ and constitutes a lever arm for the flow.

Finally, we note that the conservation of energy in a stationary
jet yields a simple relation between the terminal speed of a jet, 
and the Alfv\'en radius:
\begin{equation} 
v_{\infty} \simeq 2^{1/2}(r_A/r_o) v_K(r_o) 
\end{equation}
where $v_K(r_o)$ is the Kepler speed at the base of the field
line that threads the disk at radius $r_o$.  By combining the
conservation laws, one may show that (eg. PP92, 
\citet{Spruit96});
\begin{equation}
r_A/r_o = [(B^2/2 \pi \rho v_p \Omega_o r_o ]^{1/3} \equiv \mu^{-1/3}
\end{equation}
The dimensionless parameter $\mu$ - which is not a constant
along a given field line - has been used 
as different measure of the mass
load of a jet by \citet{Anderson05}.
 
In general, the conservation laws  
show that the mass load plays a fundamental 
role in determining both the collimation 
as well as the rotational profile of the jet.
{\it Even though direct communication of the jet with the accretion disk
is cut off beyond the fast magnetosonic surface of the outflow (which
is rather close to the disk) - nevertheless the mass load function
is transported from the accretion disk to every point in the outflow. 
This is how the accretion disk can control
the jet's dynamics even far from the source}. 

\subsection{Jet rotation: angular momentum extraction from the disk}   

Let us now carry out our second application
of the angular momentum equation (9), this time    
to find the torque that is exerted upon a thin accretion disk 
by an external field
${\bf B}$.  Any vertical flow in the 
thin disk is negligible so only the $v_r$ contribution matters,
and the rotation speed $v_{\phi}$ is 
the Kepler speed if disks are thin. 
After vertically integrating the equation,  
noting that the disk accretion
rate is $\dot M_a = - 2 \pi \Sigma v_r r_o$.  The result is the
angular momentum equation for the accretion disk under the
action of an external magnetic torque; 

\begin{equation}
\dot M_a { d (r_o v_o) \over dr_o} = - r_o^2 B_{\phi} B_z \vert_{r_o, H}  
\end{equation}
\noindent
This result shows that angular momentum is extracted
out of disks threaded by magnetic fields.  
By solving for $rB_{\phi} = k (r v_{\phi} - l)$  
and relations 
(7) and (11) for $k$ and $l$ 
the disk angular momentum equation (14) can 
be cast into its most fundamental form;
\begin{equation}
\dot M_a {d( \Omega_o r_o^2) \over dr_o} = {d \dot M_w \over d r_o}
                   \Omega_o r_A^2
                   . (1 - (r_o/r_A)^2)    
\end{equation}
As is now well understood, this fundamental
equation reveals the crucial link between the 
mass outflow in the wind, and the mass accretion rate 
through the disk; 
\begin{equation}
\dot M_a = (r_A/ r_o)^2 \dot M_w.  
\end{equation}

The magnetic field forces gas to co-rotate with the
underlying disk out to a distance of the order $r_A$
which is why these winds can be so efficient 
at extracting the angular momentum of the disk.
Later in the paper we will show that the  
typical Alfv\'en lever arm values  
in the simulations is $r_A / r_o \simeq 2-3 $
for a broad range of models. 
The mean value of the  lever arm 
allows one to calculate the ratio
of the wind mass outflow rate
compared to the disk accretion
rate,  $\dot M_w / \dot M_a \simeq 0.1 - 0.3$.

\subsection{ Two possible solutions for jet collimation: 
cylindrically collimated vs "wide-angle" flows} 

In the standard picture of hydromagnetic winds,   
collimation of an outflow occurs
because of the increasing
toroidal magnetic field in the flow as one moves
through its various critical points.  
From equation (8) one
deduces that at the Alfv\'en surface
$B_{\phi} \simeq B_p$, while in the far field
(assuming that $r_j >> r_A$) this ratio 
is of the order
$B_{\phi} / B_p \simeq r/ r_A$.  
Therefore, collimation can in principle be achieved 
by the tension force associated with the toroidal field
which leads to a radially inwards directed component
of the Lorentz force (or "z-pinch"); 
$ F_{Lorentz, r} \simeq J_z B_{\phi}$.    

The radial structure of the outflow
is found from the condition of force balance perpendicular
to the field lines in the flow,   
which is known as the Grad-Shafranov equation.  
This is a very complicated, 
non-linear equation  and no general solutions are known 
(see review, \citet{Heyvaerts03}).   
Because of the mathematical difficulties,  
analytic studies have been dominated by simplified approaches
which focus mainly on finding stationary, self-similar solutions
of the flow 
(eg. BP82, \citet{Tsinganos90,Li95,Konigl89,Ostriker98,Casse00}),     
or on analyzing the asymptotic limits of the general equations 
(HN89, PP92).  

In HN89 it was shown  
 that two types of solution are possible depending upon
the asymptotic behaviour of the total current intensity in the jet; 
\begin{equation}
I = 
2 \pi \int_0^r J_z(r',z')dr' = (c/2)
r B_{\phi} 
\end{equation}
where 
${\bf
J} = (c/4 \pi) {\nabla \times B}$ is the current density.  
In the limit that $I \rightarrow 0$ as
$r \rightarrow \infty $, the field lines are paraboloids
which fill space.  On the other hand, if the current
is finite in this limit, then the flow is collimated to cylinders.   
The collimation of a jet therefore depends upon
its current distribution - and hence on the radial distribution 
of its toroidal field - which, as we saw earlier,
depends on the mass load.  Mass loading therefore determines
the asymptotic behaviour of jets! 

In order to provide a context for the simulation results to come, it is  
useful to briefly overview
the self-similar model of BP82 in which the disk
accretion rate treated as a constant.
The only velocity in the self-similar problem is the 
Kepler speed of the disk.  Therefore, one  
anticipates that the  
various velocities in the problem scale as 
$v_A \propto c_s \propto v_r \propto v_{\infty} \propto v_K$
where one has the  Alfv\'en, sound, radial inflow,
terminal wind speed, and Kepler speed all being proportional
to one another.  
Now, the hydrostatic balance condition for  
thin disk gives $H(r) / r = c_s/v_K$.  
Therefore,  the self-similar scaling 
$c_s \propto v_K$ implies that $ H \propto r$;  
the disk is wedge-like in its spatial structure.  
This also implies that the disk temperature follows the virial 
scaling $T \propto r^{-1}$.  The density scaling
for the disk follows from the definition of the disk
accretion rate,   
$\dot M_a = 2 \pi ( 2 H \rho) v_r. r$, which
given the scaling of the radial inflow velocity  
$v_r \propto v_K$ and  
the disk scale height relation implies  
that $\rho \propto r^{-3/2}$.  The wind mass loss rate
in the wind follows since the mass loss per unit area
along the surface of the disk is $d\dot M_w / dA = \rho_o v_{(w,o)}
\propto r_o^{-2}$, which implies    
that $M_w (r_o) \propto ln r_o$.  

Note that the scaling of
the disk Alfv\'en speed $v_A \propto v_K$, together with
the density implies a particular 
scaling of the disk poloidal field, $B_p \propto r^{-5/4}$. 
In the BP82 model, the 
toroidal field on the disk should therefore scale as
$B_{\phi} / B_p = const$, so that $B_{\phi}  \propto r^{-5/4}$.
This presents a bit of a dilemma however because
the lowest energy toroidal field would be expected to
be a force-free condition which has a different scaling  
$B_{\phi}
\propto r^{-1}$.  The Alfv\'en surface in this model is
then $v_{\infty} / v_K \propto r_A/ r_o = const $; ie, a cone.  
Finally, given the scaling for the disk
magnetic field, we can now calculate
the mass load of the wind in the BP82 model;  
\begin{equation}
k_{o,BP82} \propto r_o^{-3/4}.   
\end{equation}

Jets need not be self-similar 
structures however.  In stationary states,
it is perhaps more natural to think
of them as having attained minimum energy configurations instead.  
We have seen above that for
the magnetic configuration in which the toroidal field
scales as  
$ B_{\phi} \propto r_o^{-1}$, the current intensity is finite
everywhere.  The BP82 self-similar solution leads to diverging
currents which corresponds to a higher energy state for the magnetic
configuration.   The disk-wind solutions elucidated 
by PP92 were designed to 
characterize non self-similar, minimum energy configurations.  

We briefly remind the reader of the motivation for the PP92 
solution by first writing 
down the general Grad-Shafranov (GS) equation 
for stationary, axisymmetric flow,    
for the surfaces of magnetic flux $a(r,z)$ in the flow.
Magnetic field lines are restricted to 
surfaces of constant magnetic flux,  $a(r,z) = const $, where the  
the poloidal magnetic field follows from the relation ${\bf B_p} = (1/r) 
{\bf \nabla}(a) {\bf \times \hat\phi}$.  The GS equation
can be written in a suggestive form (see PP92): 
\begin{equation}
{r^4 \over m^2} ( \nabla . {m^2 - 1 \over r^2} \nabla a) 
                   = \lambda(a) \bar \eta(a) a
\end{equation}
where the function $\bar \eta(a) \simeq const$ and the function 
\begin{equation}
\lambda (a) \equiv {\Omega_o^2 r_A^2 \over v_A^2} c_A = 
                    {4 \pi \rho_A(a)\Omega_o^2 r_A^6 \over a^2}  
                      \left ( {  d ln r_A \over d ln a} \right )^2 
\end{equation}
is a constant on a surface of constant magnetic flux and depends
upon the density at the Alfv\'en point, $\rho_A$.   
Solving this last relation for the Alfv\'en radius, we see that
$r_A \propto \lambda^{1/6}$ which is such a weak dependence that 
solutions with $\lambda = const$ were sought (PP92).

The accretion disk imposes an 
important set of boundary conditions on the GS equation 
that includes the value of the magnetic flux function across its surface; 
$a(r,z) = a_o(r_o, z_o)$.  Without loss of generality, one may assume that
the flux takes on a power-law form on the disk; 
\begin{equation}
a_o \propto r_o^{\mu + 1}.
\end{equation}
so that the poloidal field threading the disk, 
which is derived from this flux, 
becomes
\begin{equation}
B_z(r_o, 0) = br_o^{\mu - 1}.
\end{equation}

The general form of the GS equation given above can be exploited by
using the Ansatz that $\lambda = const$, as shown   
by PP92.  In this regime, we find many of the interesting
solutions including the self-similar case of BP82.
The GS equations show
that the Alfv\'en surface 
takes the form: 
\begin{equation}
r_A / r_o \propto a^{[4(\mu + 1) - 3]/[2(\mu + 1)]} 
\propto r_o^{2 \mu + (1/2)} 
\end{equation}
while the current takes the form
\begin{equation}
I(r,z) \propto a^{[1 - 2(\mu + 1)]/[2(\mu + 1)]} \propto r_o^{-\mu - (1/2)}. 
\end{equation}   

This result for the radial current distribution contains the key
insight about the collimation of jets from accretion disks.  
We note that there are two distinct categories of mass loads
that by the HN89 theorem, will have different collimation properties:
{\it For magnetic field distributions with
$-1/2 > \mu$; the radial current profile diverges}, 
\begin{equation} 
lim_{r \rightarrow \infty}I(r) \rightarrow \infty; 
\end{equation}
so that wide-angle flows are to be expected.  On the other 
hand, {\it distributions with $ -1/2 < \mu$ have currents 
that vanish at infinite radius,}
\begin{equation}
lim_{r \rightarrow \infty}I(r) \rightarrow 0 
\end{equation}
so that the flow should collimate to cylinders. 

The particular state investigated by
PP92 is the unique case in which the current is finite, but does
not diverge;
$\mu = -1/2 $, 
whose radial current profile is  
\begin{equation}
I_{PP92} = const, 
\end{equation}
and poloidal field
profile in the disk is $B_p \propto r_o^{-3/2}$.
The Alfv\'en surface for the PP92 model scales as  
$(r_A/r_o)_{PP92} \propto r_o^{-1/2}$,   
while the terminal speed in this solution has
the behaviour $v_{\infty} \propto
r_o^{-1}$.  
Finally, the mass loss rate in the wind for 
the PP92 solution is $\dot M_w \propto
[(r_o/r_i) - 1]$, where $r_i$ is the inner edge of the disk.
The 
mass loading of this model wind scales as 
\begin{equation}
k_{o,PP92} \propto r_o^{-1/2}.   
\end{equation}

How does an accretion disk establish the
radial profile of the outflow mass load function,  
$k(r_o)$?  The density at the  
disk surface is prescribed by the hydrostatic balance condition with
the base of the disk corona above it (the initial
condition we have used in our own simulations, see \S 4).  
In reality, we expect that the jet
originates from some region in the disk corona so that the
principle of hydrostatic balance between the disk and the base 
of the corona should remain reasonably secure.  The second
factor in the mass load is the injection speed of material into
the base of the outflow, 
$v_{\rm inj}$.  This speed must be subsonic in magnitude.   
We consider that the most natural scaling of this injection speed
is with the local Keplerian velocity at each radius
of the disk; typically $v_{\rm inj} = 10^{-3} v_K$ at any point
on the disk. 
These two physically reasonable conditions imply that the main factor
in determining the mass loading of the disk wind is actually the 
radial distribution
of the threading magnetic field across the disk.
Assuming a power-law scaling of this poloidal field component
given above, we find that the mass load function at the disk surface
follows a power law relation;
\begin{equation}
k(r_o) = \rho_o v_{p,o}/B_{p,o} \propto 
 \propto r_o^{-1-\mu} 
\label{keqn}
\end{equation}

How do the mass loads of the different wind models
compare?  For the BP82 and PP92 mass-load functions as an
example:
the PP92 mass loading function 
is a less steeply raked function of
disk radius than the BP82 self-similar case 
(wherein $ k_{BP82} \propto r_o^{-3/4} $). 
The associated current
profile for the BP82 solution 
(which can also be found from equations (19) and (20)
above (with $\lambda = const$) and pertains to 
the case $\mu = -1/4$. 
The current profile in the BP82 theory is therefore 
\begin{equation}
I_{BP82} \propto r_o^{- 1/4}. 
\end{equation}

\begin{figure*}
        \centering
        \includegraphics[width=0.9\textwidth]{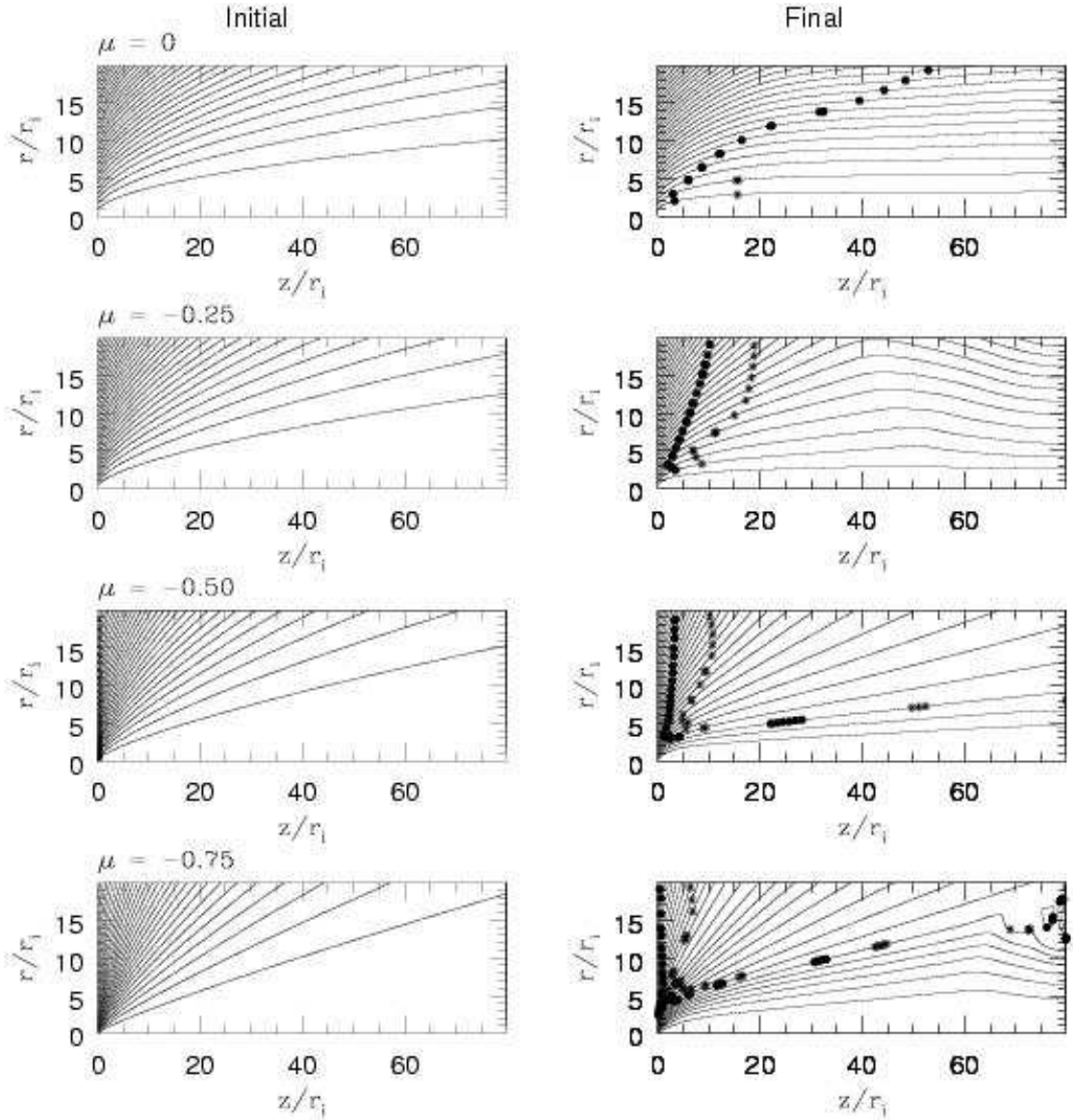}
	\caption[]{Left panels: initial magnetic field configurations
         for winds with initial potential ($\mu = 0$), Blandford
         \& Payne ($\mu = -0.25$), Pelletier \& Pudritz ($\mu = -0.5$),
          and 7Q ($\mu = -0.75$) configurations shown.  Right panels:
          final magnetic field configurations (at t = 400)
          for each case, with Alfv\'en
          points (filled circles) and FM points (stars) marked.
          Note the more open magnetic - and 
          stream line structures as the initial magnetic configuration
          (value of $\mu$) changes. }
        \label{fig1}
\end{figure*}

The essential point 
is that whereas the BP82 current $I(r)_{BP82} \rightarrow 0$ as
$r \rightarrow \infty$,   
the Pelletier-Pudritz has a finite
current in this limit $I(r)_{PP92} \neq 0$.
This implies that 
yet steeper mass load profiles  
and more generally, the monopole-like magnetic configuration
of models like \citet{Romanova97}, and the X-wind of \citet{Shu00},
should diverge in this limit.   The mass load of these
latter models produces a  
toroidal magnetic field that is strongly concentrated towards the 
outflow axis, with little left in the wide-reaches of the outflow
to effectively collimate the outflow to the axis.  

Thus, we have shown that by 
using the HN89 limit together with the theory 
of mass loading shown above, that the PP92 case
should provide a dividing line between jets that ultimate collimate
to cylinders, and jets whose magnetic and streamlines are parabolic
and therefore "wide-angled".  
This result implies that disk winds can in principle provide a wide-angle
pressure that can drive molecular outflows, and that this is essentially 
determined by how the accretion disk mass-loads the jet.

\section{ Numerical Simulations - initial conditions}

Our numerical approach in this paper is a generalization of 
the method used in 
our earlier papers (OPS97, OP97a,b; OP99). 
Conditions on the underlying accretion disk
remains constant in time and provide a
set of fixed boundary conditions for
the outflow simulation.
Several groups have exploited this approach (eg. 
 \citet{Ustyugova98,Romanova97,Meier97,Krasnopolsky99,Fendt02}). 
The published simulations differ
in their assumed initial conditions, such as the
magnetic field distribution on the disk,
the plasma $\beta$ ($ \equiv P_{\rm gas} / P_{\rm mag}$)
above the disk surfaces, the state of the initial
disk corona, and the handling of
the gravity of the central star.

We used the ZEUS 2-D code   
which is arguably the best documented and utilized MHD code
in the literature (\citet{Stone92a,Stone92b}).    
It is an explicit, finite difference code that runs on a
staggered grid.  The evolution of the magnetic field is 
followed by the method of constrained transport.  
In this approach, if ${\bf \nabla . B} = 0$ holds for the
initial magnetic configuration, then it remains so for all
later times to machine accuracy.  The obvious way of securing
this condition is to use an initial vector potential ${\bf A(r,z,t=0)}$ 
that describes the desired initial magnetic field at every point in the 
computational domain.

The accretion disk in all of our work so far is chosen to be  
initially surrounded by a polytropic corona ($\gamma = 5/3$) 
that is in hydrostatic balance with
the central object.  We do this because it is the 
equation of state used in BP82 
which provides an invaluable analytic and numerical
solution to which we can compare our simulations.
The accretion disk at the
base of the corona is given a density profile
that it maintains at all times in the simulation, since the disk
boundary conditions are applied to the "ghost zones"
and are not part of the computational domain.
It is chosen so that the disk surface is  
in pressure balance with the corona above it
(ie. $\rho \propto r_o^{-3/2}$ for a $\gamma = 5/3$ 
model - as in BP82).
This hydrostatic state has a simple analytic solution
which was used as the initial state for all of our
simulations.  

\onecolumn
\begin{figure}
    \subfigure[Radial dependence of the mass loading parameter, $k$ at the disk surface.  The first two cases fit the predicted
behaviour; $k \propto r^{-1-\mu}$.  (A power law form $\psi = br^a$ is the
 fit used for any physical quantity $\psi$).]
    {
      \label{fig2a}\includegraphics[width=.5\textwidth]{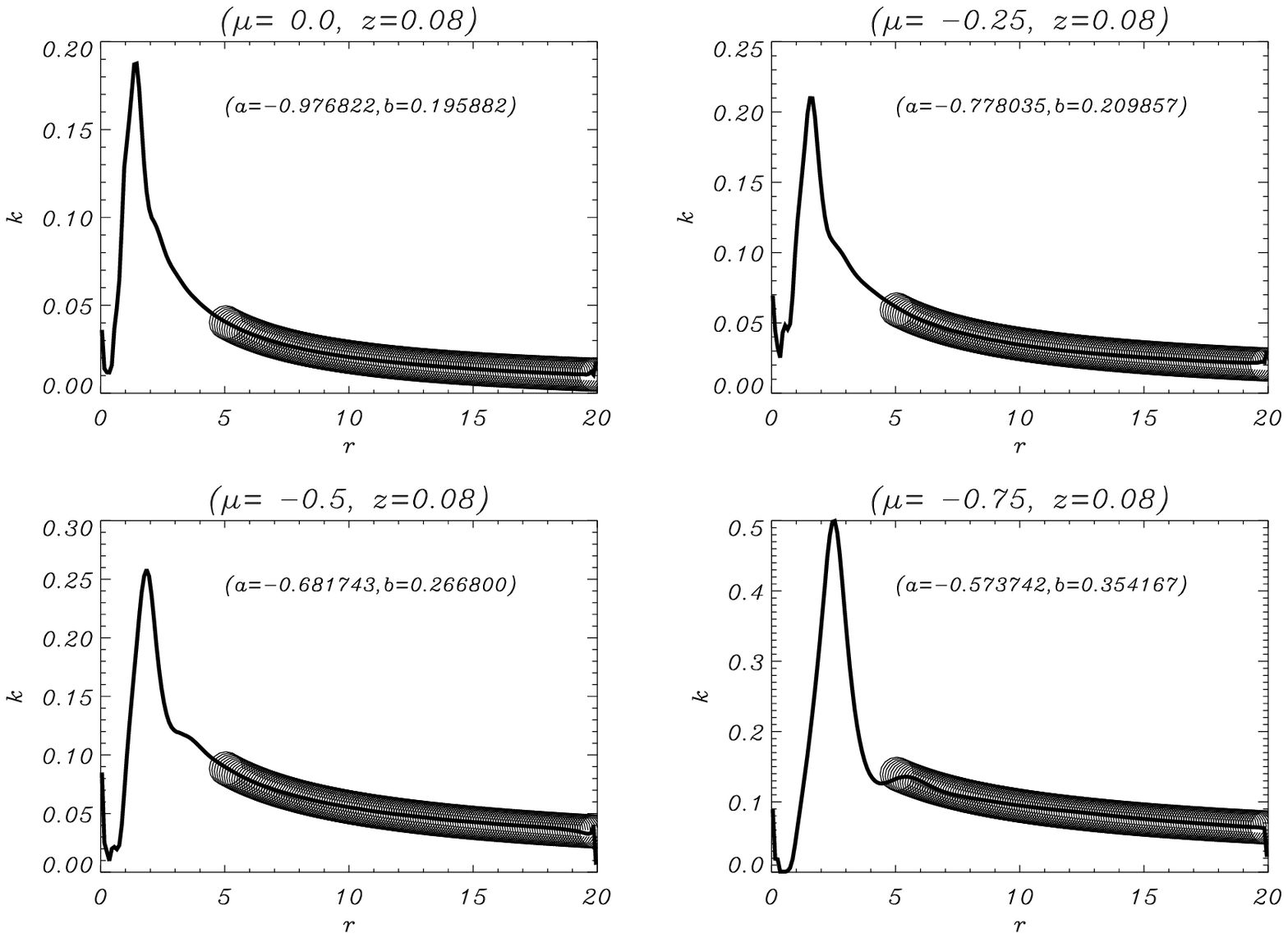}
    }
    \subfigure[Radial dependence of density for each configuration at high $z$.
           The peak density (which defines the jet) in the first
          two cases is about an order of magnitude greater than latter two, and
is collimated closer to the disk axis.]
    {
      \label{fig2b}\includegraphics[width=.5\textwidth]{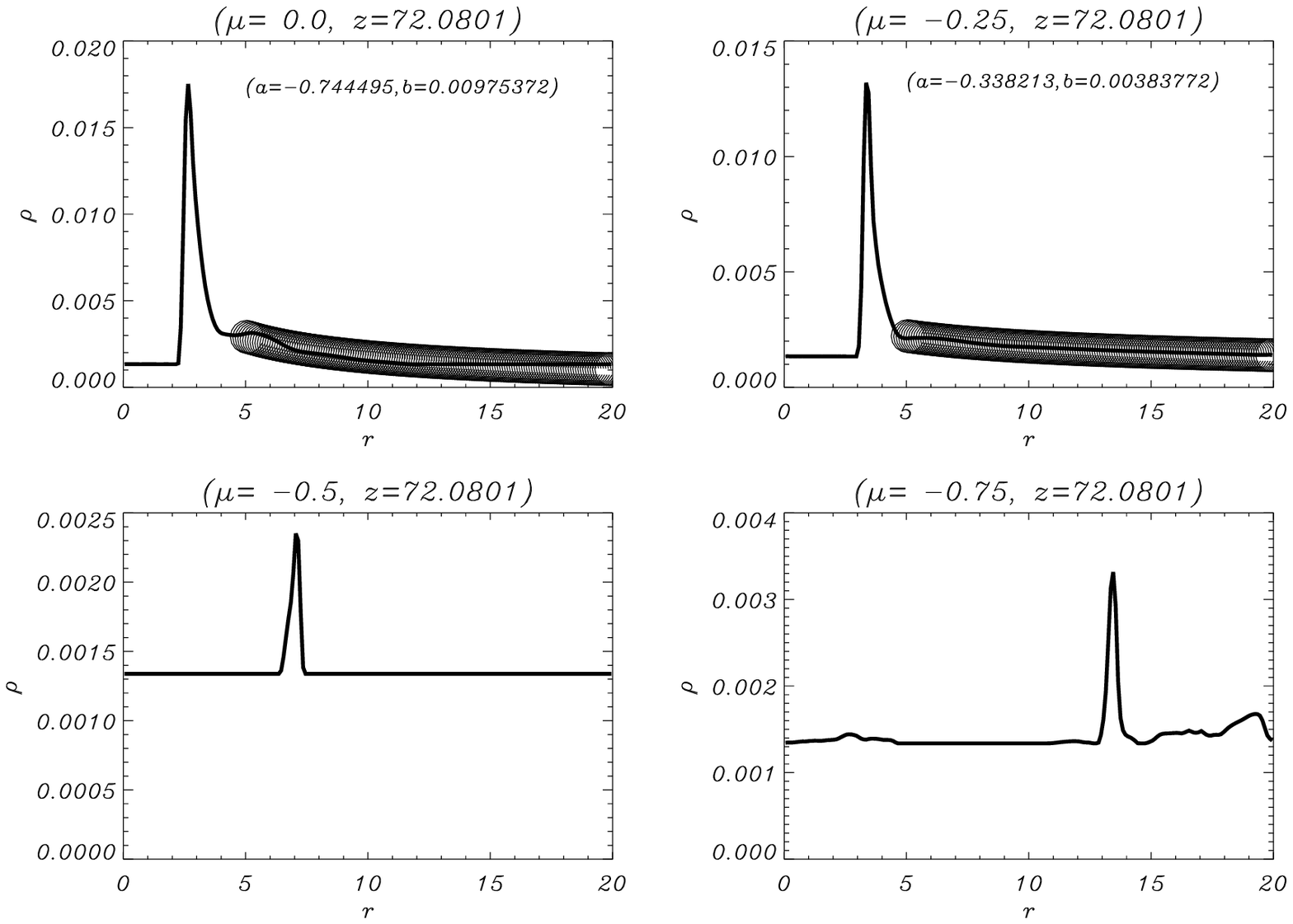}
    }
    \subfigure[Radial dependence of the poloidal velocity at high $z$.
        The collimated flows show higher velocities closer to the disk axis, whereas the non-collimated solutions have almost the opposite behaviour,
        indicative of a wide-angle wind.]
    {
      \label{fig2c}\includegraphics[width=.5\textwidth]{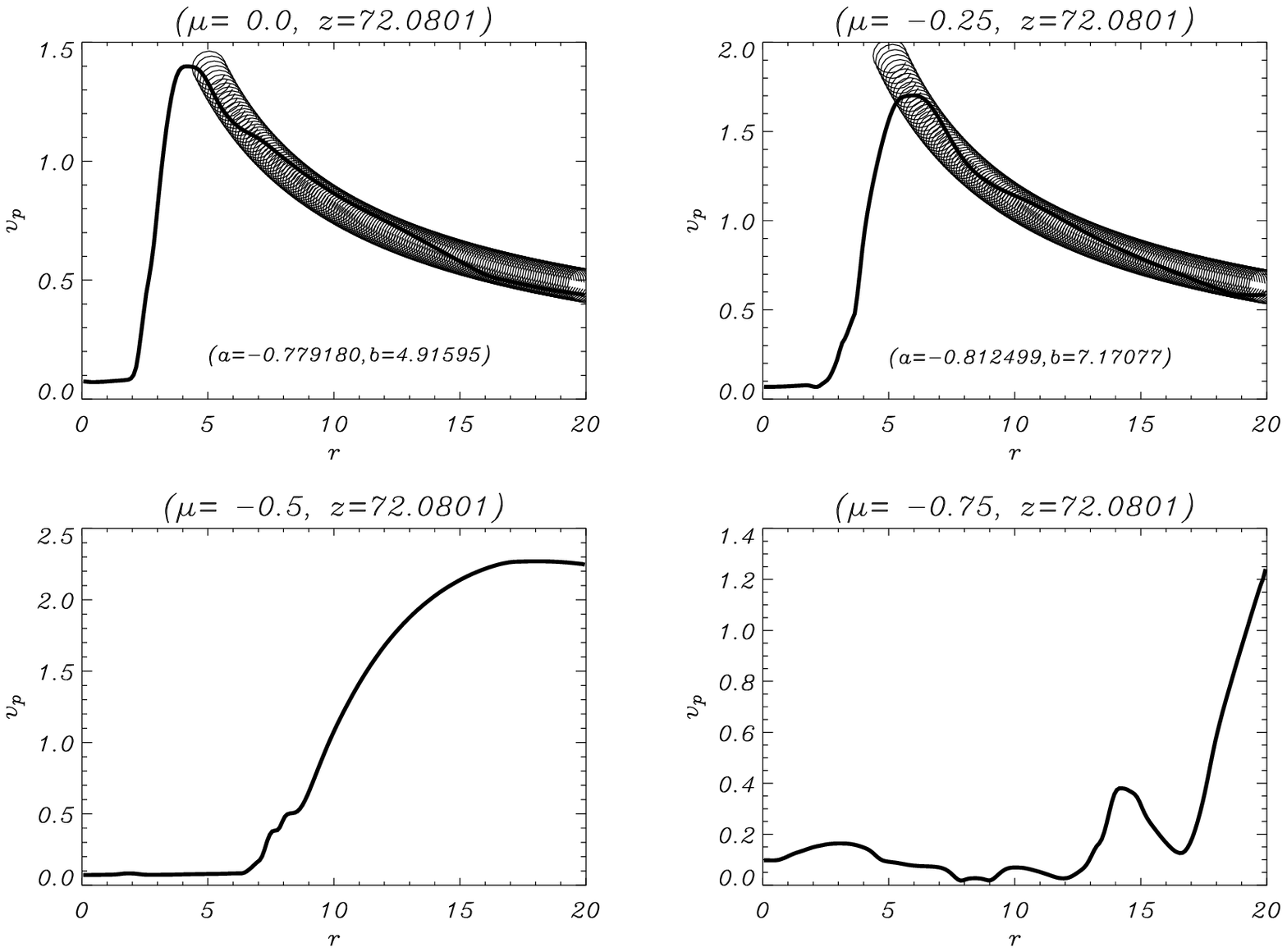}
    }
    \subfigure[Radial dependence of the toroidal magnetic field for each configuration at high $z$. The maximum value is close to the disk axis in the first two cases, as expected for a collimated wind.  The last two cases 
 exhibit a much weaker field strength, explaining why these flows show no collimation.]
    {
      \label{fig2d}\includegraphics[width=.5\textwidth]{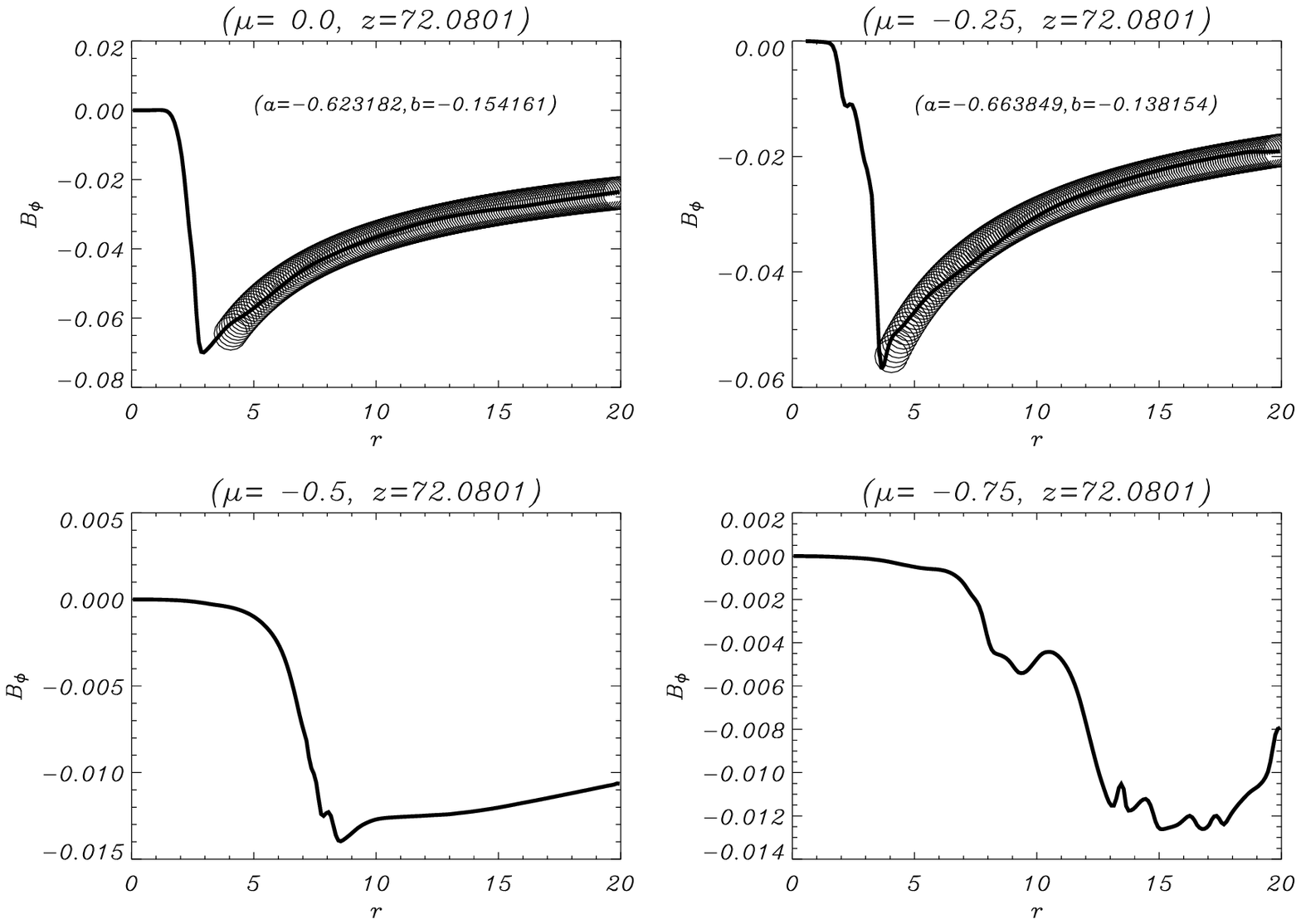}
    }
   \caption{Radial profiles of physical quantities at $t=400$.
 The cuts shown in th upper left panel, (a), were taken at $z\simeq 0.0$ close
 to the disk surface  while the cuts in panels (b), (c) and (d) were taken further out at $z\simeq 72.0$.}
   \label{fig2}
\end{figure}
\twocolumn

%\twocolumn

\begin{figure}
        \centering
        \includegraphics[width=.5\textwidth]{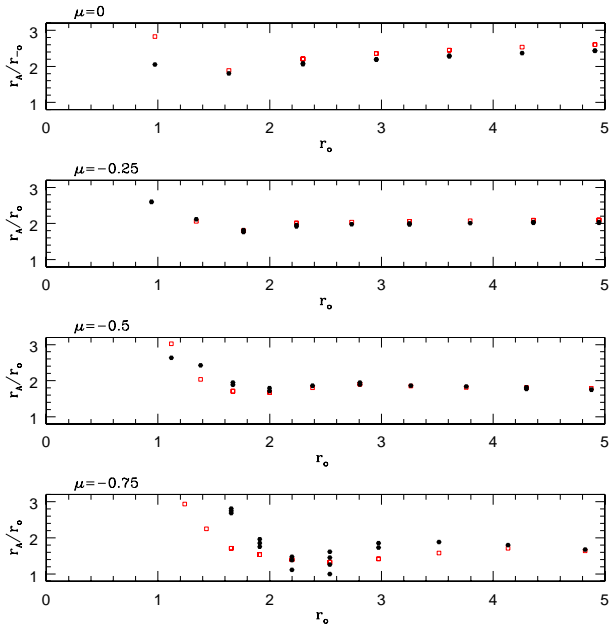}
        \caption[]{The Alfv\'en lever arm for the four 
	configurations, at $t=400$, as compared 
	to the prediction from steady-state theory (squares).} 
        \label{fig3}
\end{figure}

The initial magnetic configurations in our 
studies are chosen so that no
Lorentz force is exerted on the initial 
(non-rotating) hydrostatic corona described above;      
more specifically, we used initial current-free
configurations ${\bf J } = 0$ in the computational 
domain.  
The initial vector potential is subject to the boundary
condition on the disk that the poloidal field on the
disk surface $(r_o, 0)$ has a power-law form and 
field structure given by equations (23) and (24) respectively. 

An analytic solution for the field throughout the 
computational volume can be found for the choice 
$\mu = 0$ on the disk.  
This is the so-called 
"potential" configuration that we used in several of our papers.  
We can accommodate more general  
initial magnetic configurations however, by using 
a Hankel transform technique to solve for
the initial vector potential that is
subject to these constraints (see Appendices A and B for details).  
We chose our particular Hankel transform method as the one that  
best maintained the initial equilibrium state
for arbitrary amounts of machine time (see Appendix B). 
We always use an unsoftened gravitational
potential from the central object. 
The initial corona is ultimately swept away by the disk wind in all of 
these simulations so that the mass source for our jets 
becomes dominated by the material injected from the 
fixed disk below.    

We simulated a representative set of models  
by choosing an initial
BP82 model ($\mu = -1/4$),
PP92 ($\mu = -1/2$), and a yet more steeply declining
magnetic field such as $\mu = -3/4$ (referred to as the 7Q configuration).  
These initial configurations are plotted out in the left panels of fig.~\ref{fig1}.  
From the current profile given in equation (24), we would
predict that if these flows settle into a stationary state, then
the models with $\mu = 0; -1/4$ should be cylindrically collimated
while the remaining two models; $\mu =-1/2;-3/4$ should be "wide-angle"
or parabolically collimated flows.  This was our rationale in choosing
these special, illustrative models. 
The mass loading functions corresponding to these four magnetic 
field distributions are progressively less steeply raked; going
from: 
\begin{equation}
k(r_o) \propto r_o^{-1}, \quad r_o^{-3/4}, \quad r_o^{-1/2}, \quad 
r_o^{-1/4}
\end{equation}
respectively.  

Our simulations of thin disks require that we specify five physical 
quantities at all points of the disk surface 
at all times.  There are the disk density $\rho(r_o)$; 
components of the vertical and toroidal magnetic field, 
$B_z(r_o)$ and $B_{\phi}(r_o)$; and velocity components in the 
disk, $v_z(r_o)$ and $v_{\phi}(r_o)$.  The remaining field
component $B_r(r_o)$ is determined by the solenoidal 
condition while the radial inflow speed through the disk
$v_r(r_o) \simeq 0$ for the purposes of the simulations since
it is far smaller than the sound speed in a real disk. 
The model is described by five parameters, whose
values describe conditions at the inner edge of the disk
at radius $r_o = r_{\rm i}$. Three of these parameters describe
the initial corona: $\beta_{\rm i} $ which is 
typically 1/3 (and which therefore falls with radius);  
the density jump across the corona/disk boundary
$\eta_i$ which is typically 100; and 
ratio of the Keplerian to thermal energy density $\delta_i$, set to be 300
initially. Two additional parameters describe the disk physics.  
The parameter, $\nu_i$,   
which scales the toroidal field in the disk $B_{\phi} = \nu_i/r_o$, 
was found to be unimportant.  We  
emphasize that this last initial condition plays no role in the
subsequent evolution of the simulation - even if the initial
toroidal field on the disk surface vanishes, the wind itself
quickly generates the self-consistent field.  We have never
observed that this parameter is important. 

We ran high resolution simulations in which 
$ (500 \times 200)$ spatial zones were used
to resolve a physical region of $ (80 r_i \times 20 r_i ) $ in
the z and r directions, respectively.
Our simulations ran up to $400 t_{\rm i}$
(where $t_{\rm i}$ is the Kepler time for an orbit
at the inner edge of the disk, $r_{\rm i}$).
The standard parameter settings were
\begin{equation}
(\eta_i, \nu_i, v_{inj}, 
\delta_i, \beta_i) = (100.0, 1.0, 0.001, 300.0, 1/3)
\end{equation} 

We focused our simulations on the four initial magnetic field 
configurations discussed above - which between them give a very broad
range of mass load profiles and behaviours for the jets.
We found that we 
could only run the BP82 configuration to $t=400$, 
and barely that, as the simulation ground down terribly.
Once the density in the corona had been cleared out, 
the Alfv\'en velocity went very high, making the 
timestep prohibitively small.  
This was also a problem with the PP92 case.  
Since the 7Q case never really blew out a
lot of material, we managed to run it out to $t=500$.

\begin{figure}
        \subfigure[Density structure of jets from the potential ($\mu = 0$, 
         left panels), and Blandford-Payne ($\mu = -0.25$, right panels) solutions. 
	 Note the densest material moves along the jet axis.]
	{        
	\label{fig4a}\includegraphics[width=.5\textwidth]{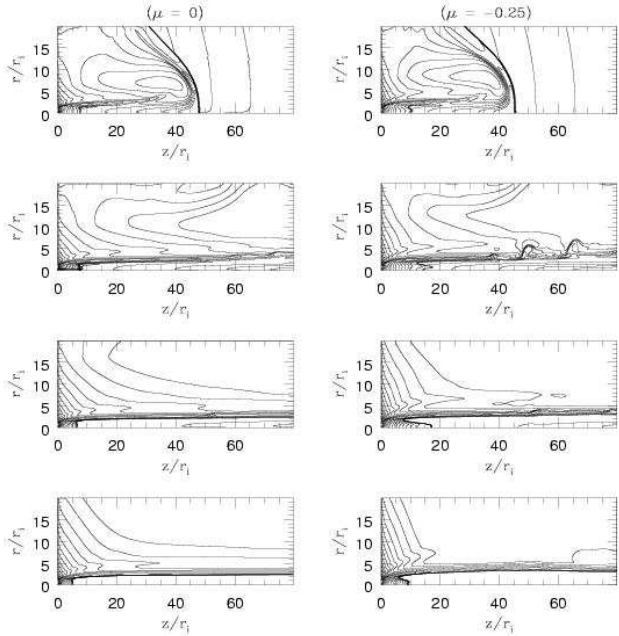}
        }
        \subfigure[ The corona has been cleared out by the uncollimated disk wind.  
	  The 'jets' from the Pelletier-Pudritz
         ($\mu = -0.5$, left panels), and steeper ($\mu = -0.75$, right panels) solutions
	should be compared to the velocity profiles in Fig.~\ref{fig3}]
	{
        \label{fig4b}\includegraphics[width=.5\textwidth]{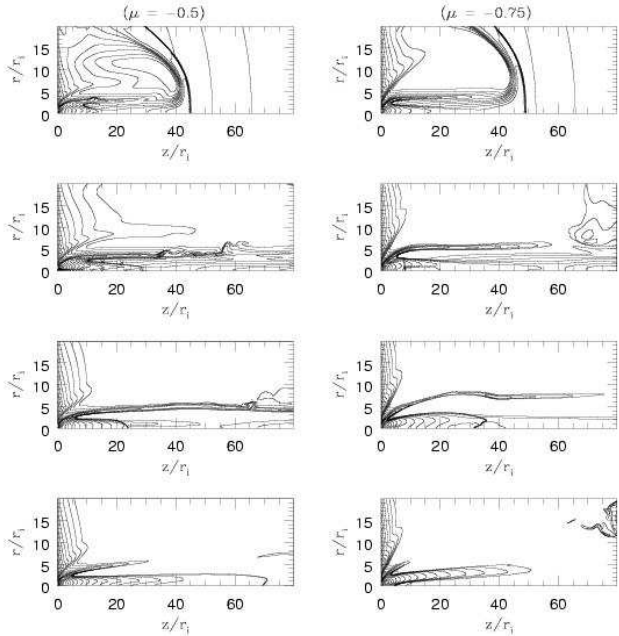}
        }
	  \caption{Contours mark surfaces of constant density 
         in the jet at four different times (t= 60, 120, 200, 400).}  
        \label{fig4}
\end{figure}

\section{Results}

\subsection{Jet collimation - magnetic field lines, density, and 
velocity behaviour}

Figure \ref{fig1} plots the magnetic field lines of both the 
initial states of each model (left panels) as well
as the final states of simulations.  In addition
to the potential case ($\mu = 0.0$), simulations of the 
Blandford-Payne ($\mu = -0.25$),
Pelletier-Pudritz ($\mu = -0.5$), 
as well as an even more steeply raked ($\mu = -0.75$)
initial magnetic configuration are shown.  
It is evident that the field lines in these
right hand panels are progressively less well collimated towards the axis, 
suggestive of a wide angle flow at larger radii.
A careful examination shows that while the potential and BP82 cases
achieve cylindrical collimation within our computational grid,
the PP92 and 7Q cases remain more "open".  While collimation may still
occur on spatial scales much larger than our computational volume, 
this is difficult to investigate because of 
the very small time step required in regions of large Alfv\'en
speeds.  We see a  
difference in collimation behaviour that agrees 
with the predictions of steady-state jet theory that we outlined
in \S 3.2.  It appears that the PP92 case is  
the transition between 
solutions that collimate to cylinders, and those that have 
parabolic structure in the asymptotic limit.

Figure 1 also shows   
the positions of the Alfv\'en surfaces for each 
configuration marked on the field lines. 
The BP82 final state shows that the Alfv\'en surface
is indeed nearly a cone as expected while the PP92 simulation qualitatively 
agrees with the expected scaling $r_A \propto r_o^{1/2}$ (see \S 3.2).
The details of the outflow for the "potential" configuration
are found in OPS97 and OP97a.  

The mass load function for these models is shown in Figure~\ref{fig2a} wherein
the values of the function $k(r,z)$ are numerically tabulated  
along a radial spatial cut through the flow that is 
a few pixels above the 
disk inside the computational zone.  
In stationary state, the mass
loads should settle down to power-law radial distributions that match
the input at the surface of the disk, predicted in 
equation (31).  The figure shows power-law fits
for physical quantities $\psi$ (in this case, $\psi = k$) of the form

\begin{equation}
\psi = br^a
\end{equation}
which give for the 4 different magnetic configurations (always starting
with the potential configuration $\mu = 0$), $a= -0.98, -0.78, -0.68,
-0.57$, which compares favourably with the predictions except
for the last case, the 7Q model.
The simulations of the potential flow and the BP82 outflows have settled 
into stationary states so that the mass loads in the jet closely give
the input value at the disk surface (-1.0 and -0.75 respectively).
The PP92 simulation has not settled down completely by the end of our
run, but well enough that there is a good correspondence with the  
predicted load profile of -0.5 compared to our numerical value -0.68.  
The greatest
discrepancy is found for the 7Q run, but as noted, this has not achieved
a completely stationary state even by the end of $t=400$.   

A quantitative analysis of the radial density behaviour of these 
4 outflows is given in Figure \ref{fig2b}. The potential and BP82 configurations
have greater peak densities than the PP92 and 7Q models.  Another 
difference appears to be that whereas the former two have a declining
power-law density behaviour with 
\begin{equation}
a=-0.74 \quad (\mu = 0); a= -0.34 \quad (\mu = -1/4)
\end{equation}
- the potential and BP82 cases respectively -  the radial behaviour is quite different
in the PP92 and 7Q cases which appear to be nearly independent of radius.
In every case, there is a central "spike" in the gas density which is the
material at the smallest radius that moves more parallel to the outflow
axis.  The radial position of this "spike" moves outward in radius as
one can see from the figures.  

The corresponding radial behaviour of the poloidal velocity is analyzed
 in Figure \ref{fig2c} by means of a spatial cut through the flow
taken most of the way down the jet at $z = 72.08$.
The peak speed in each of
the first 3 models increases.  There is also a distinct fall-off
in jet outflow speed with disk radius, which for the potential 
and BP82 cases behave as
\begin{equation}
a = -0.78, \quad a= -0.81;  
\end{equation}
which are virtually identical.  
It is evident from this figure that the PP92 and 7Q cases have 
a larger "hole" in the middle of their jets, and that their
poloidal speeds fall off less steeply with jet radius.  

The radial profile of the toroidal magnetic field 
for the models is shown in Figure \ref{fig2d}.  
In all of our simulations, the toroidal
field must vanish on the rotation axis.  It reaches a peak value at
the inner edge of the outflow and then declines.  This means that 
the toroidal magnetic field pressure exerts an inward pressure force
from this peak, and an outwards pressure gradient force at larger
radii than the peak radius (see OP97a).  While the potential and
BP82 cases have similar radial power law behaviour for the toroidal
field; $B_{\phi}(r) \propto r^{-0.64}$ on average at the larger 
radii, the PP92 and 7Q cases have much weaker toroidal fields everywhere.

%\clearpage
%\onecolumn
\begin{figure*}
    \subfigure[Potential case: Note that the highest velocities are nearest the outflow
         axis, and that the flow is very well-collimated]
    {
      \label{fig5a}\includegraphics[width=.75\textwidth]{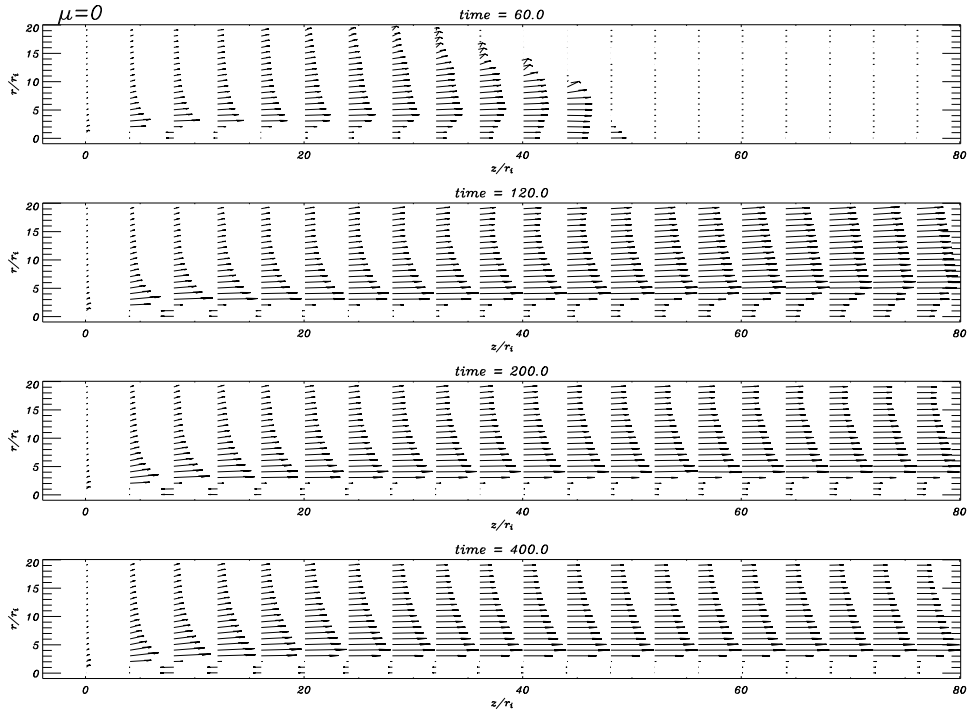}
    }
    \subfigure[BP82 case: Note that the highest velocities are nearest the outflow
         axis, and that the flow gently recollimates]
    {
      \label{fig5b}\includegraphics[width=.75\textwidth]{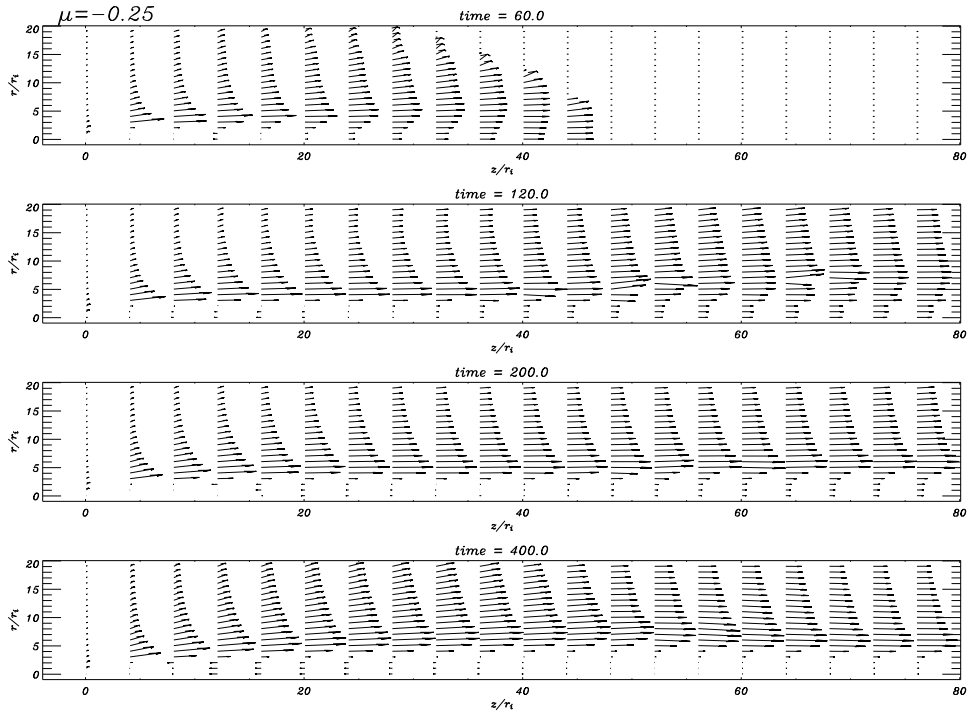}
    }

  \caption{Poloidal velocity fields of the 4 models, shown at the 4 different times in Fig.~\ref{fig2a} }
  \label{fig5}
\end{figure*} 
%\twocolumn
\addtocounter{figure}{-1}
\begin{figure*}
\addtocounter{subfigure}{+2}
     \subfigure[PP92 case: The flow seems to be collimating in the early stages of the jet evolution, but
        as can be seen in the last panel, the flow assumes a conical geometry.]
    {
      \label{fig5c}\includegraphics[width=.75\textwidth]{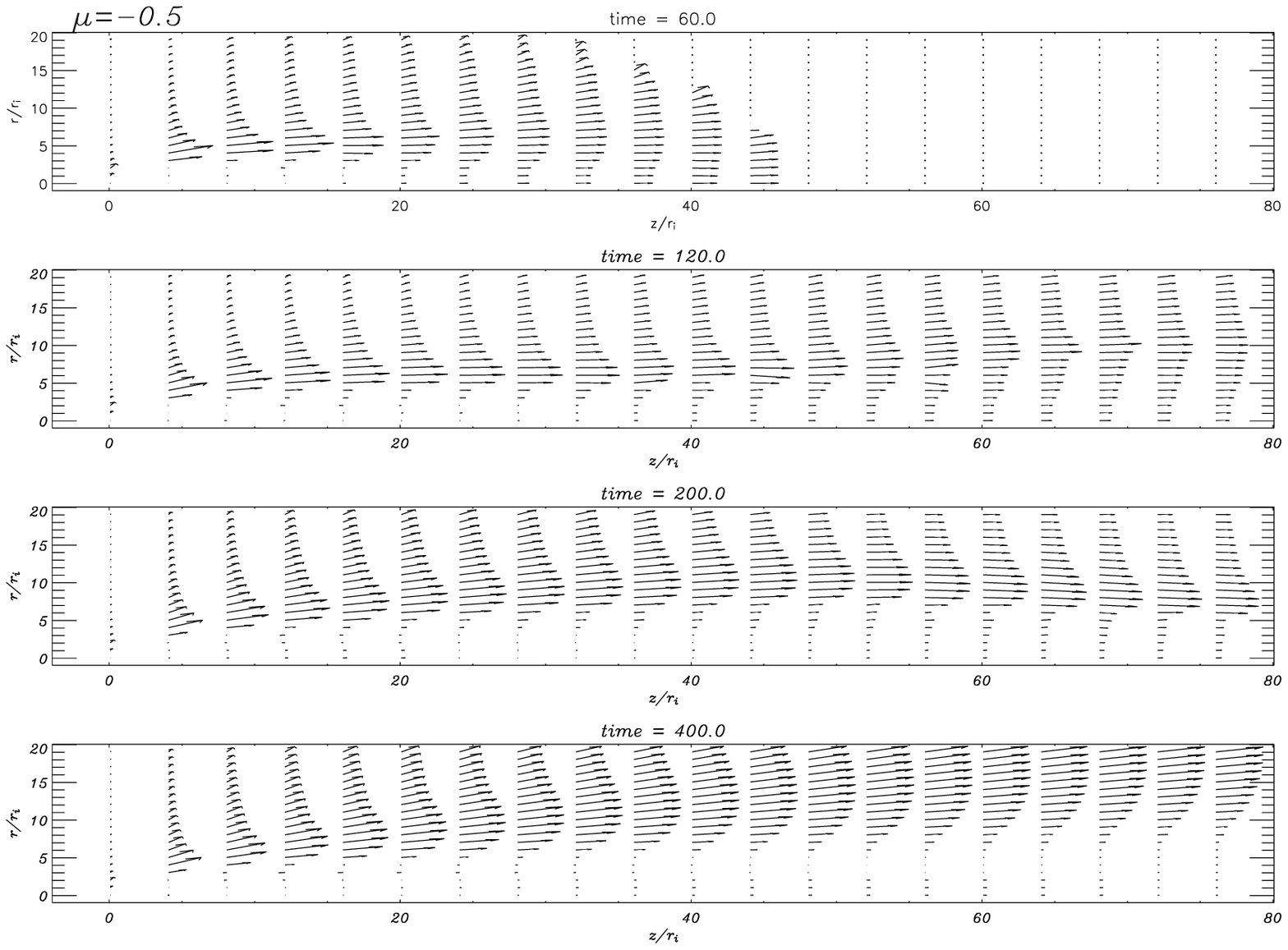}
    }
    \subfigure[7Q case: The steepest configuration barely collimates at all and evolves differently
        from the other solutions.  The wide-angle wind is even more extreme than the
        Pelletier-Pudritz solution]
    {
      \label{fig5d}\includegraphics[width=.75\textwidth]{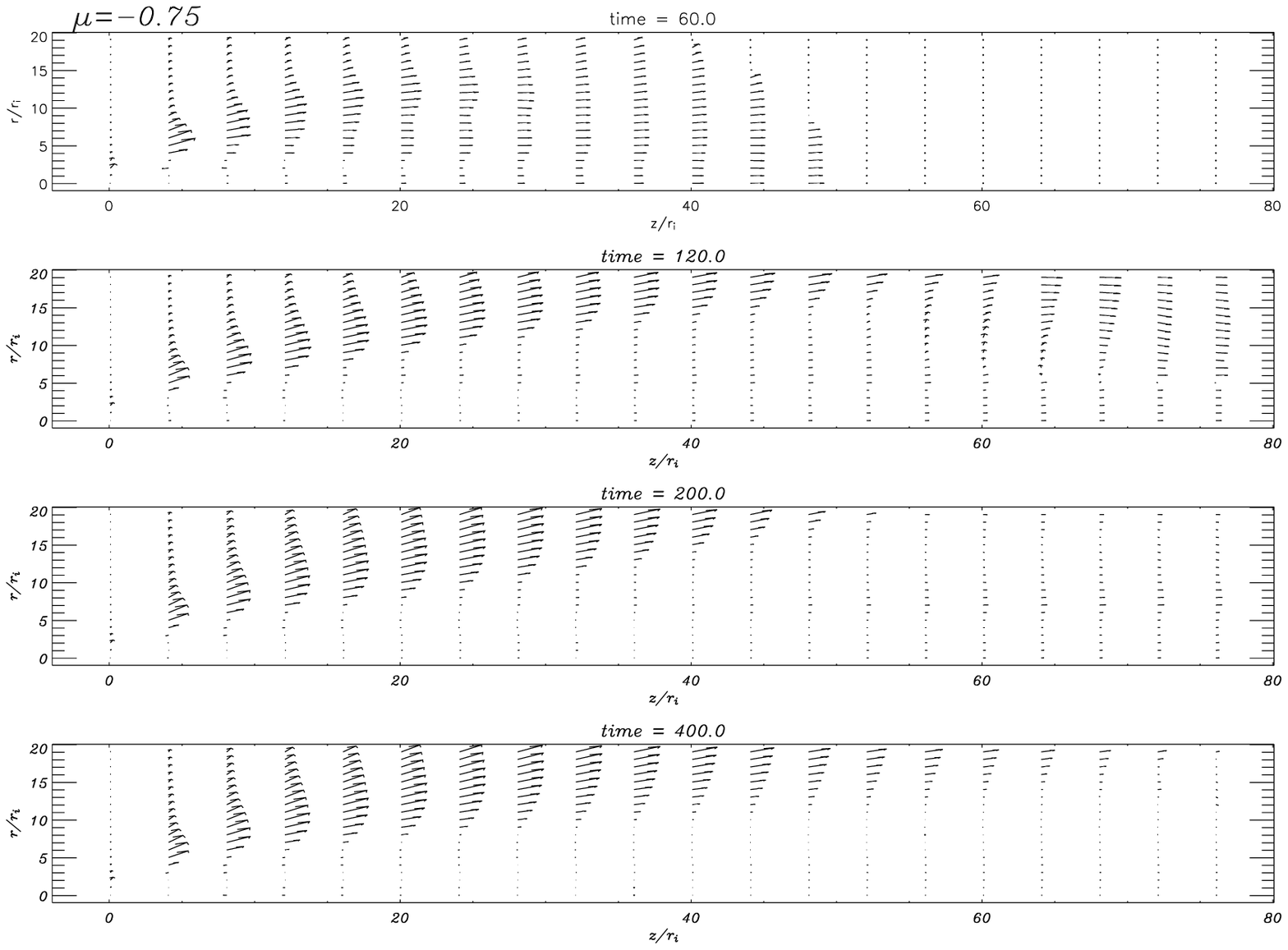}
    }
  \caption{...continued.}
\end{figure*}

We compare the predicted structure of the Alfv\'en surface
and that from theory in Figure 3 where we plot
the Alfv\'en lever arm versus the footpoint radius of a given field
line.  The analysis of equation (25) predicts that as one
goes from $\mu = 0 \rightarrow \mu = -3/4$, then 
\begin{equation}
r_A/r_o \propto
r_o^{1/2}, \quad const., \quad r_o^{-1/2}, \quad r_o^{-1}
\end{equation}
respectively.
The 
first three solutions compare well to the theory.  
The steepest case has its Alfv\'en 
surface too close to the disk surface to be well-resolved, 
which may explain the discrepancy.
We see that the Alfv\'en lever arm usually lies within the domain
of 2-3 in value for the innermost field lines of the flow.

The evolution of the density structure of these jets is shown
in the snapshots
presented in Figures \ref{fig4a}--\ref{fig4b} 
for 4 different times ($t= 60, 120, 200, 400$).  
In each simulation, we see the jet driving a 
bow shock through the quiescent gas in the first frame.  In the second 
frame the shock has just moved out of the grid.  In all four
of these simulations, one clearly sees that the densest gas
in the jet is always well collimated and moving nearly parallel  
to the axis.  This is a general consequence of any reasonable
mass loading for a jet.   Thus, whether or not the outflow
has a wide-angle aspect to it or not, the bulk of the material
that we see in all of our simulations moves along  
in a fairly concentrated and collimated manner.
Thus, all of our simulations when observed in appropriate 
forbidden line emission, 
would appear "jet-like" - as first 
noted by \citet{Shu00} for 
their X-wind simulations.

The poloidal velocity vectors associated with these outflows is shown  
in Figures~\ref{fig5a}--\ref{fig5d}.
The four frames shown in each
of these panels correspond to the 
4 times shown in Figs.~\ref{fig4a}-\ref{fig4b}
A well-collimated jet is visible in 
the velocity figures for the first two cases - potential 
and BP82 outflows.  In the fourth case (7Q) we see that whereas the jet    
density is highly collimated (Fig. 4b, right panel) along the axis -  
a wide-angle velocity profile is obvious.
Thus, the bulk of the jet material follows the few field lines that have collimated
parallel to the disk axis, 
but is clearly sub-Alfv\'enic. 
The bulk of the kinetic energy of the flow is in the wide-angle wind.  
Thus, the PP92 and 7Q cases represent two models
in which a wide-angle
component could carry momentum sufficient to produce a wide-angle
molecular outflow. 
The final stage ($t=400$) of the PP92 solution resembles 
the 7Q case  more closely than 
either of the well-collimated solutions.

Taken together, Figures \ref{fig2}--\ref{fig5} show the 
existence of two regimes of disk-driven winds; one 
that results in a highly collimated jet, 
and the other that produces a wide-angle wind.   
The PP92 configuration seems to be a transitional model 
lying between these two distinct regimes.
In all situations, the bulk of the flow's mass density is fairly
well collimated along the flow axis.  The cylindrical or 
parabolic collimations 
($\mu = 0,-1/4$ and $\mu = -1/2,
-3/4$ cases respectively) 
primarily affect the velocity field and
a small amount of the gas in the outflow.  

\subsection{Jet dynamics - moving along a field line}

\begin{figure}
    \subfigure[Potential case]
    {
      \label{fig6a}\includegraphics[width=.5\textwidth]{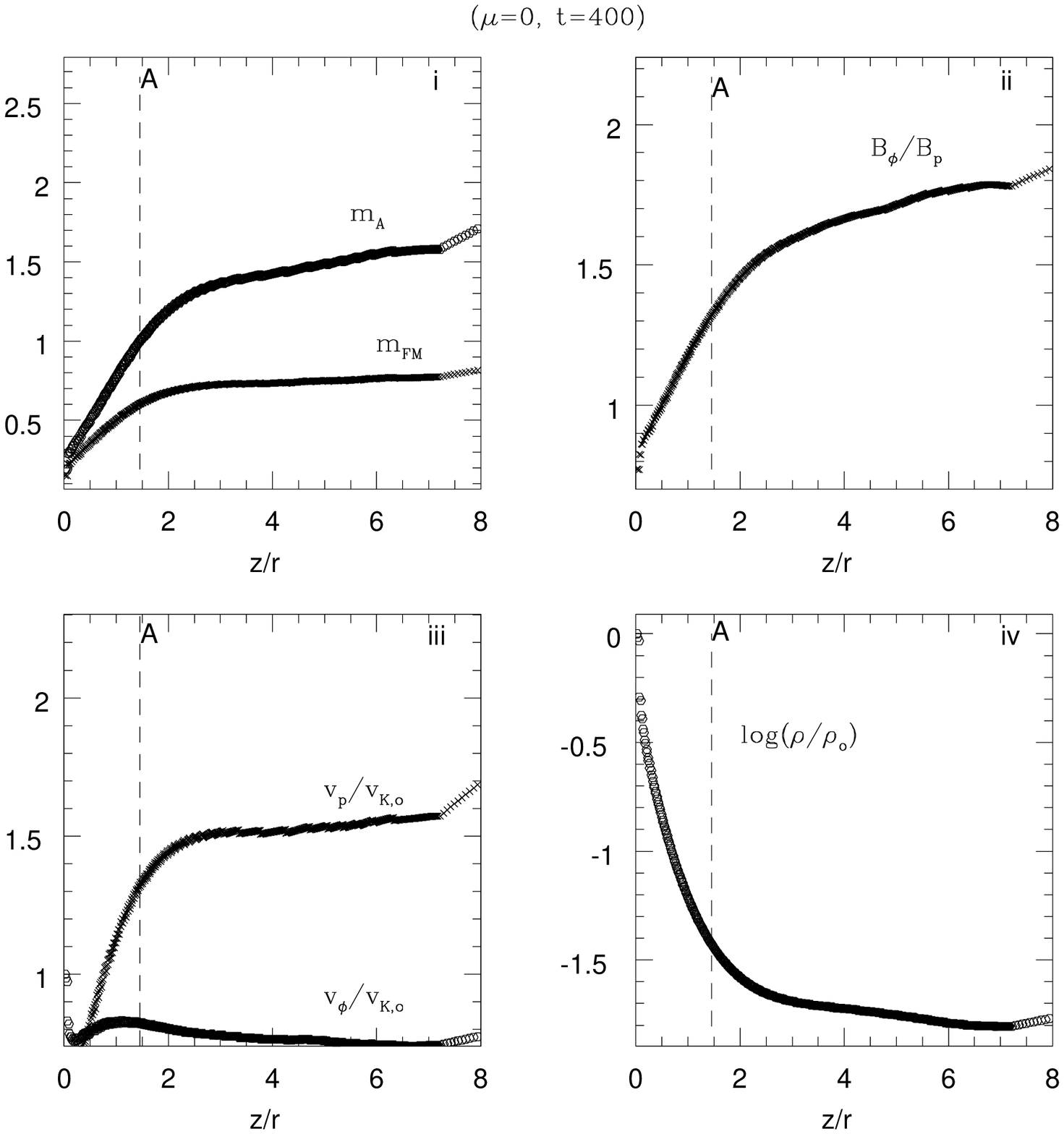}
    }
    \subfigure[BP82 case]
    {
      \label{fig6b}\includegraphics[width=.5\textwidth]{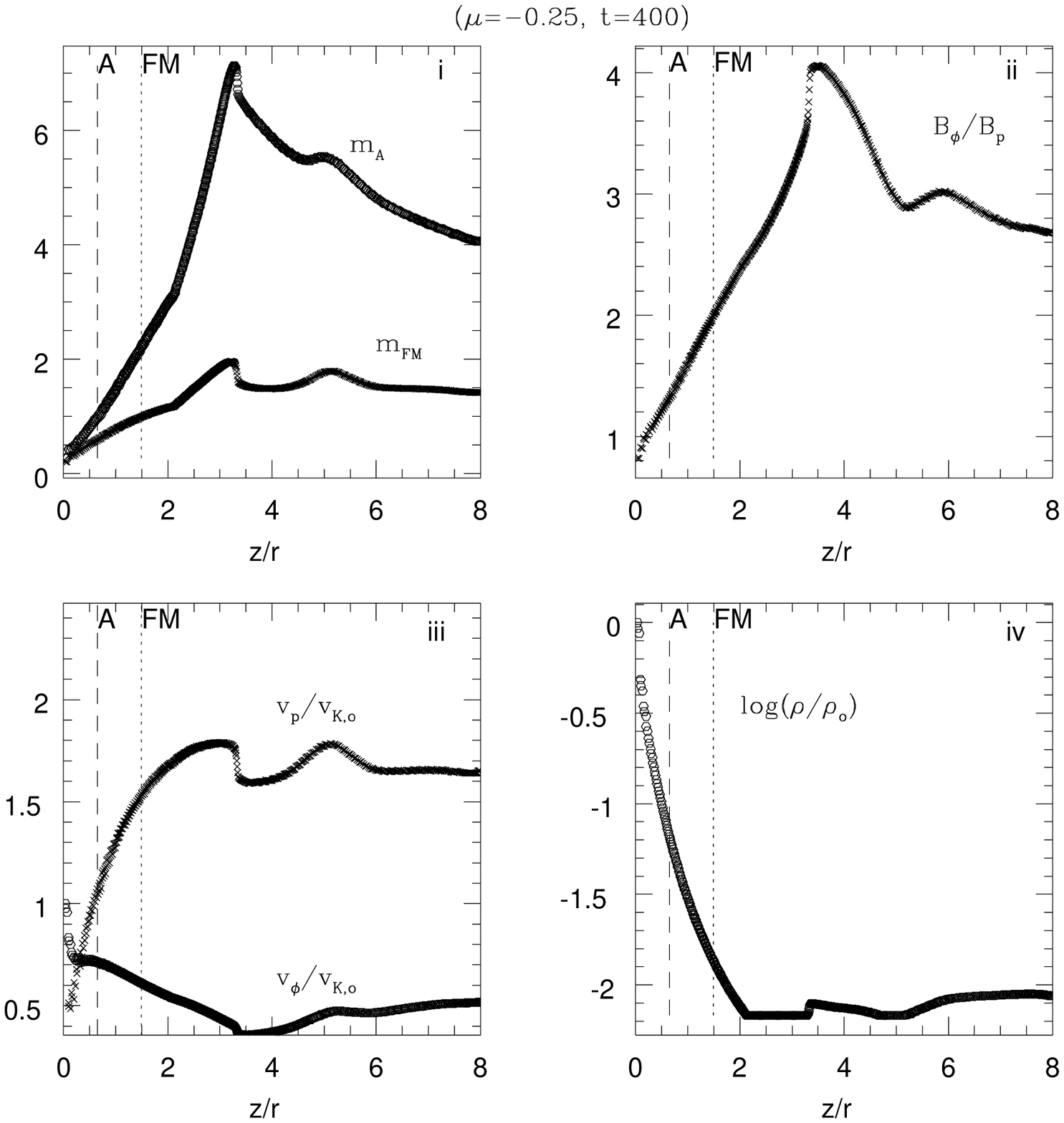}
    }
    \subfigure[PP92 case]
    {
      \label{fig6c}\includegraphics[width=.5\textwidth]{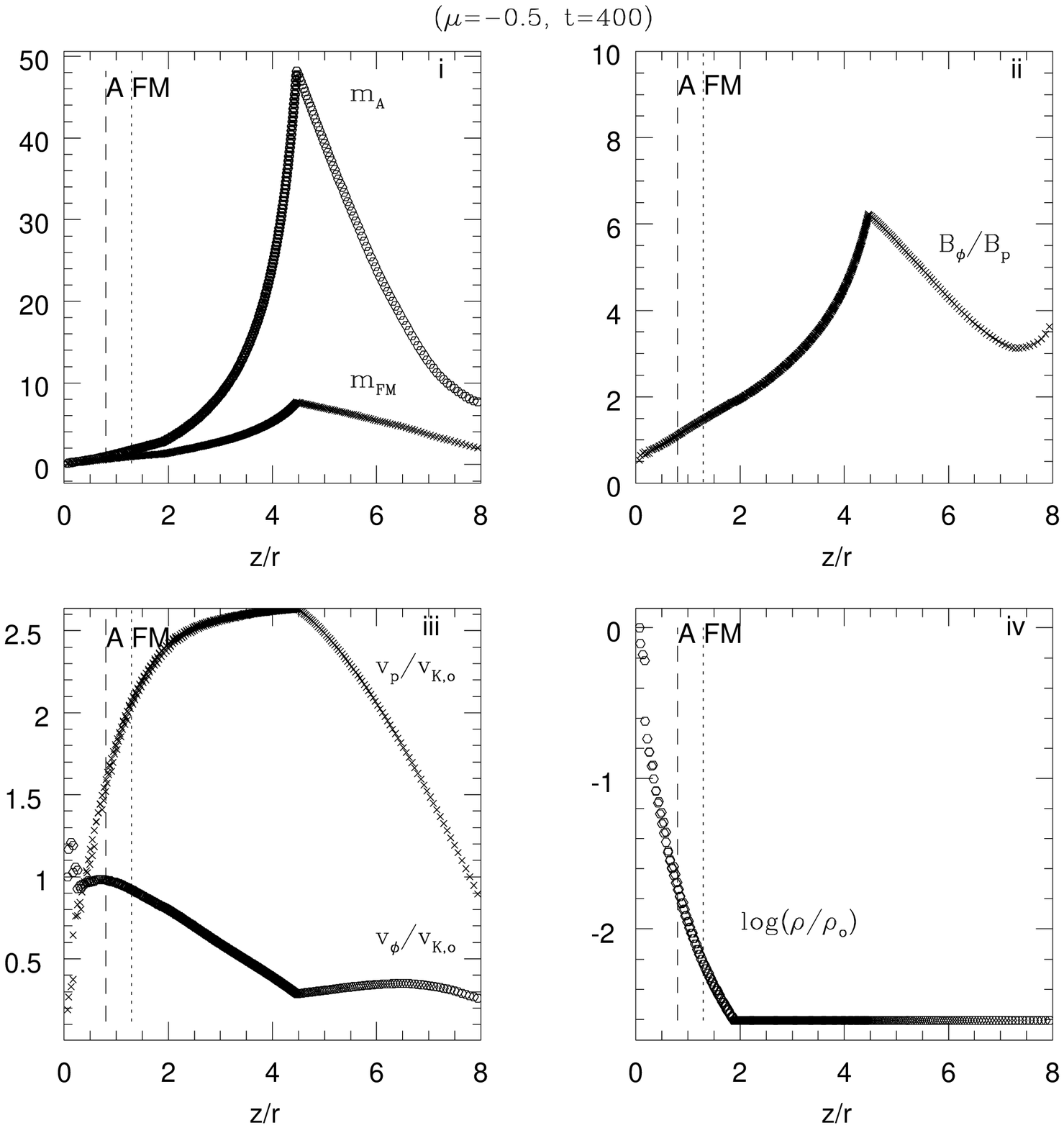}
    }
    \subfigure[7Q case]
    {
      \label{fig6d}\includegraphics[width=.5\textwidth]{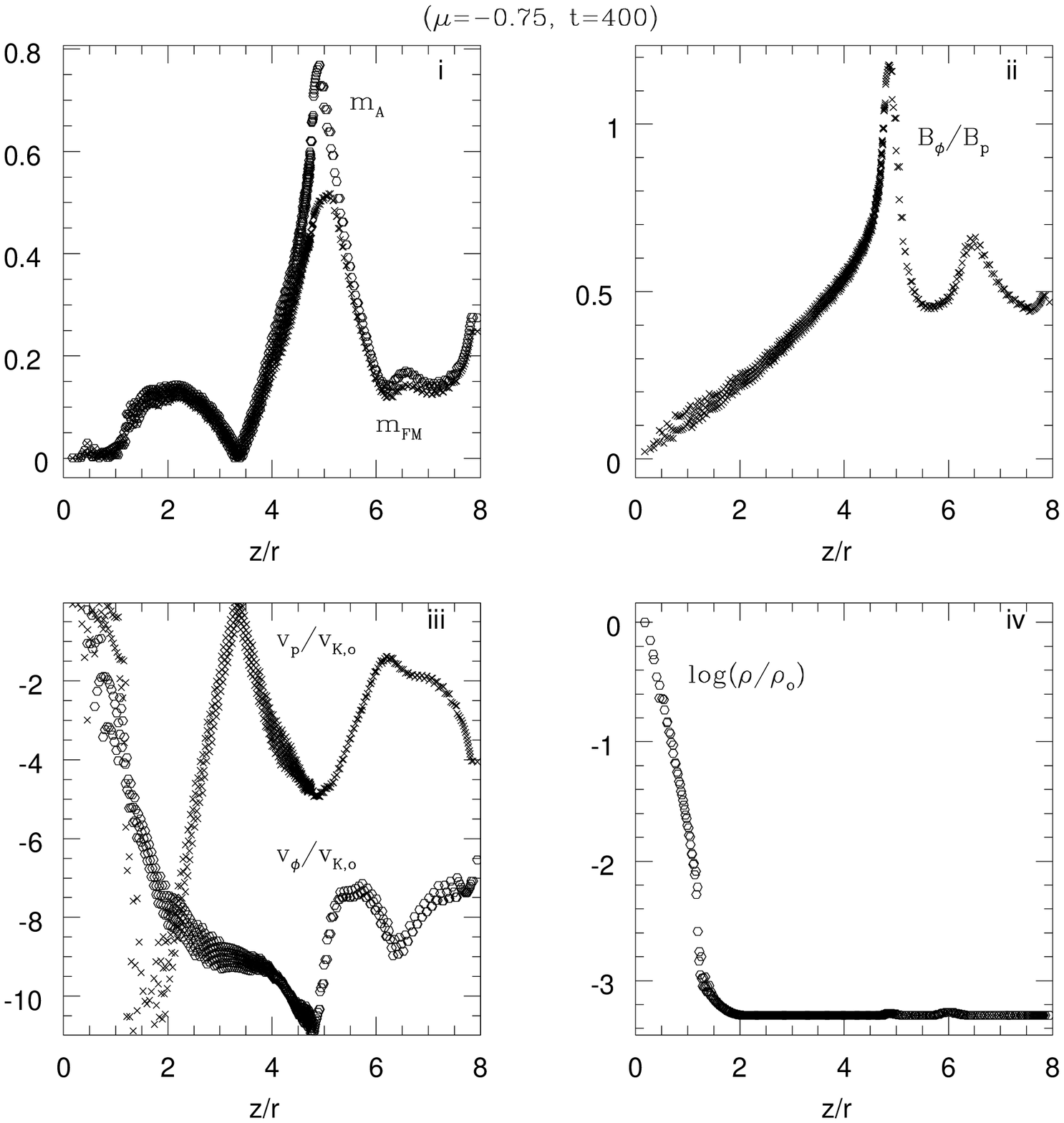}
    }

  \caption{Physical quantities along the fifth innermost 
	  fieldline, plotted against z/r.  The position of the Alfv\'en and fast magnetosonic
	  (marked by FM) critical points are shown.}
  \label{fig6}
\end{figure}

In Figs.~\ref{fig6a}$--$\ref{fig6d}, we trace the behaviour of physical quantities
along the fifth innermost fieldline which is chosen since it
remains on the computational
mesh and is concentrated more towards the central axis of the flow.  
Comparing the upper-left panel for each configuration, 
we see that the potential case reaches an 
Alfv\'en Mach number of about 1.5, 
but never reaches 
its fast magnetosonic speed.
The BP82 case accelerates much more quickly, 
and has reached its fast magnetosonic point at $z=1.5$, 
reaching a peak Alfv\'en Mach number of 7 before the 
poloidal velocity levels off.  

The PP92 configuration reaches a very high Alfv\'en Mach 
number of nearly 50 before it undergoes an abrupt downturn
in value.  The 7Q case does not really seem to settle into
as stationary a configuration, but the results do seem to suggest
that it is in a different regime of Alfv\'en Mach number.

The rather abrupt decrease in the outflow Mach numbers 
seen in this figure arises
from the presence of shocks and knots along the innermost
field lines.  This is evident from an examination of the
Alfv\'en and FM points along inner field lines shown in 
Figure \ref{fig1}.  While the potential flow seems to have reached
a very stable, stationary state by the end of the simulation,
the BP82 and PP92  simulations require increasingly longer 
times to settle into stationary flows while the 7Q case
did not settle into a stationary state in the far
field region at $z/r_i \ge 60$.  
Another indication that shocks are responsible for these
abrupt changes is that  
the toroidal to poloidal magnetic field, as well as the value
of the toroidal flow velocity, also undergo similar sharp changes in 
value.

The outflow velocity along the field line, in units of the
Kepler speed at the foot of the field line, increases for these
first three models.   Thus the highest speed flow
in the centre-most region of a jet arises in the PP92 model,
more even than the BP82 solution. As for the  
rotation speed of material in the
outflow, it always falls to a fraction of its initial value, as
theory predicts (because of the expansion of 
the flow, as well as the transport of angular momentum that is
carried by the twisted field).  
We defer our analysis to the radial rotational profile of 
material in the jet to \S 5.4 below. 

\onecolumn
\twocolumn
%\clearpage

%\clearpage
\onecolumn
\begin{figure}
    \subfigure[Potential case]
    {
      \label{fig7a}\includegraphics[width=.5\textwidth]{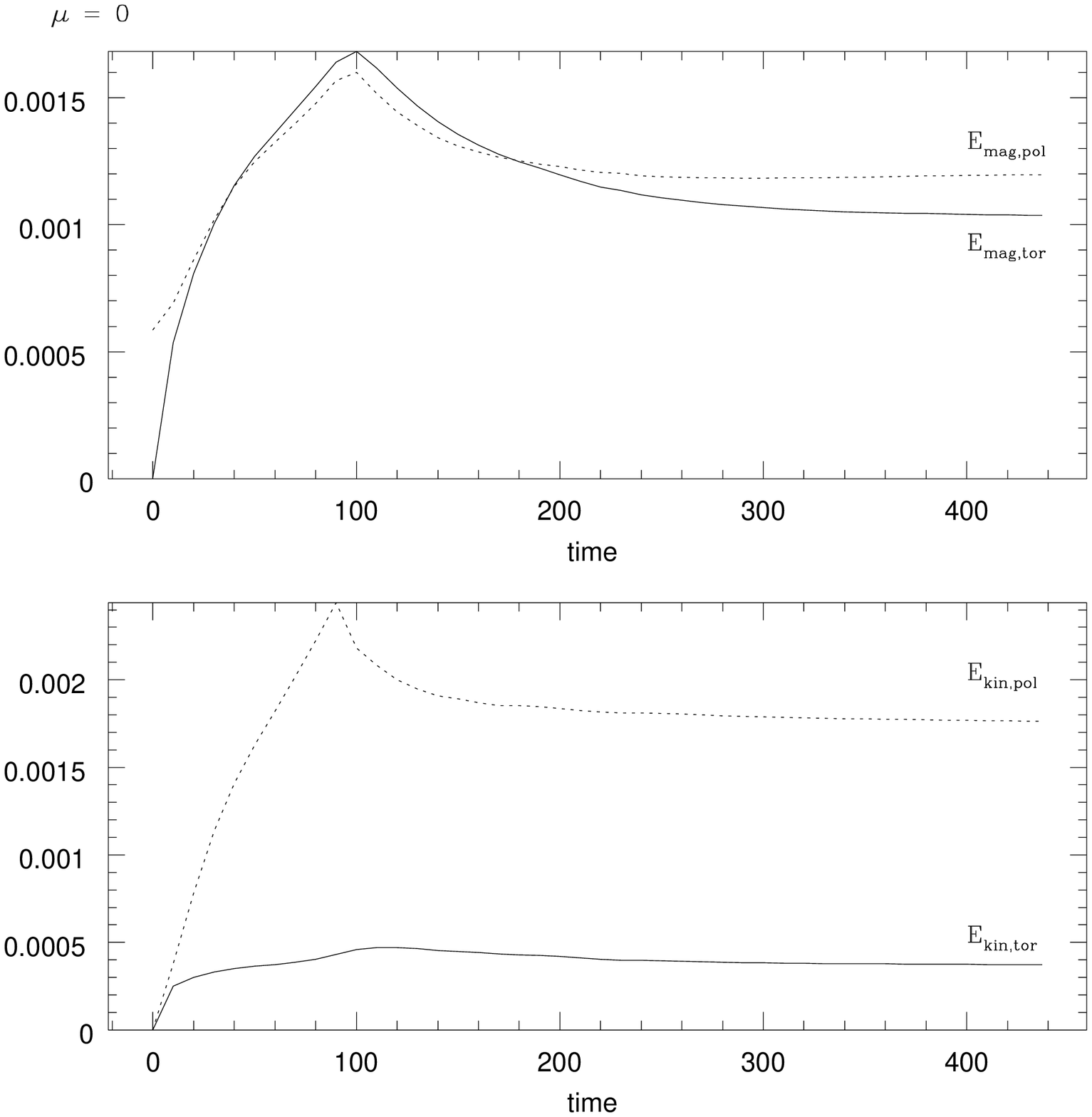}
    }
    \subfigure[BP82 case]
    {
      \label{fig7b}\includegraphics[width=.5\textwidth]{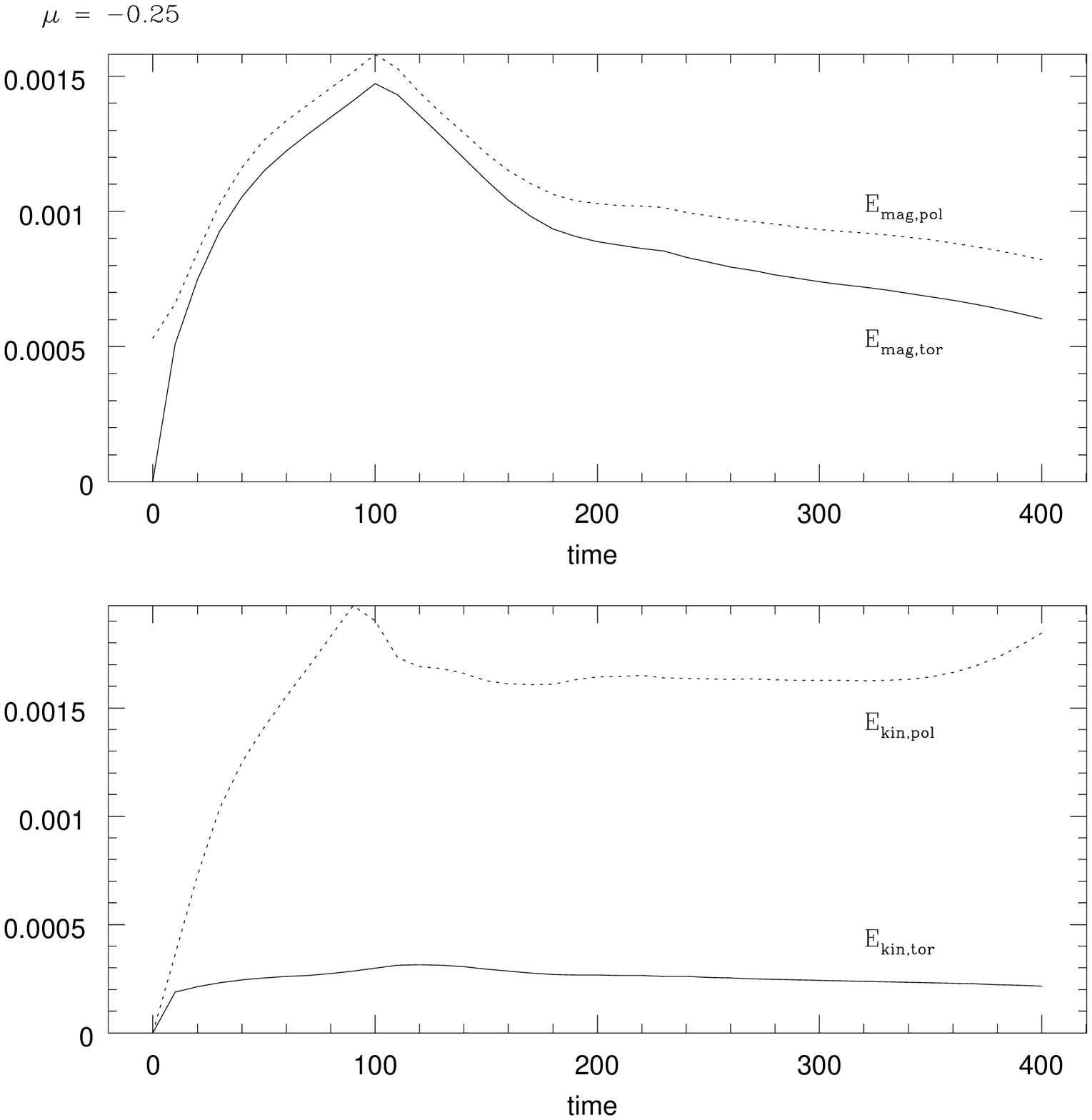}
    }
    \subfigure[PP92 case]
    {
      \label{fig7c}\includegraphics[width=.5\textwidth]{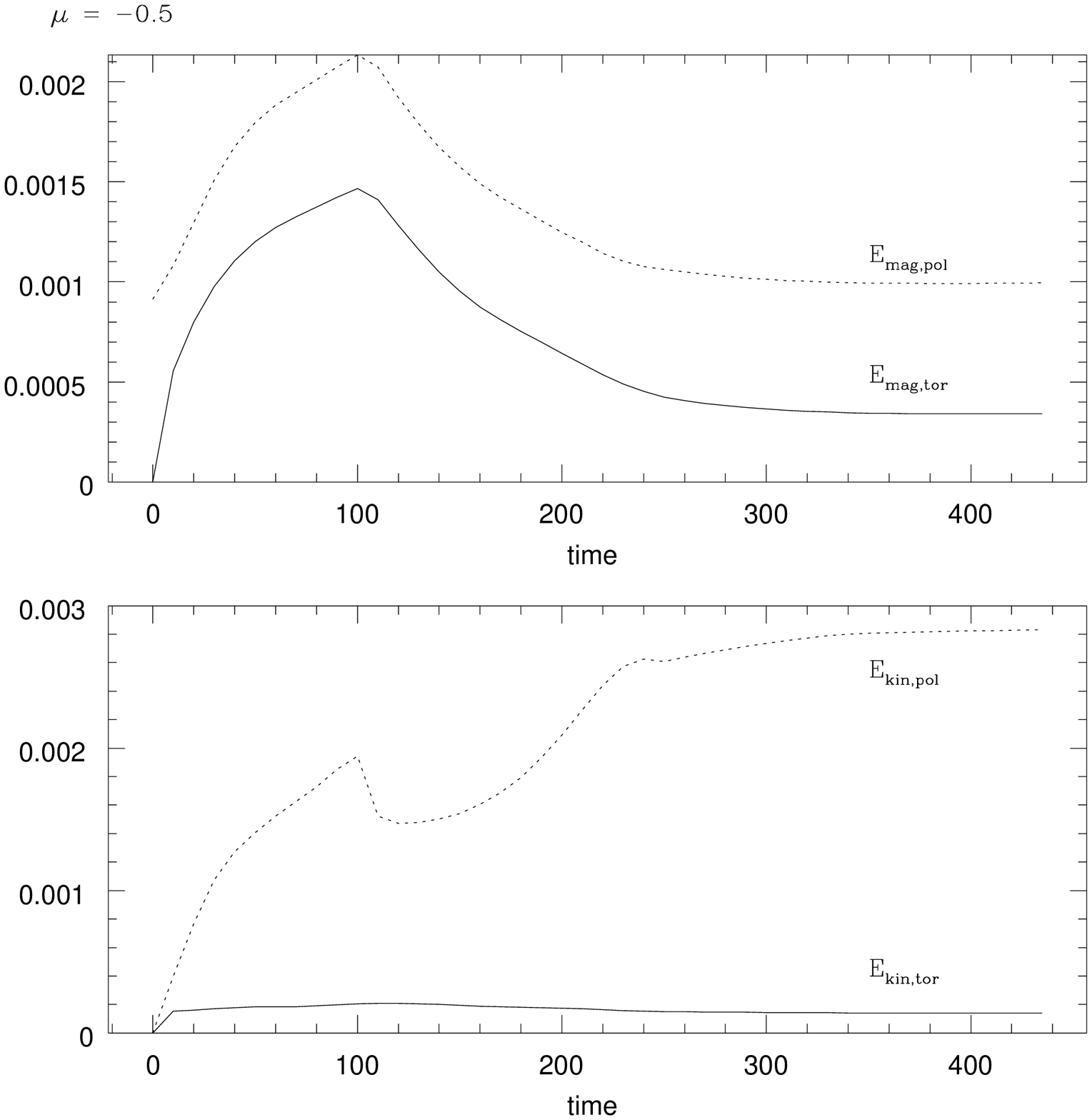}
    }
    \subfigure[7Q case]
    {
      \label{fig7d}\includegraphics[width=.5\textwidth]{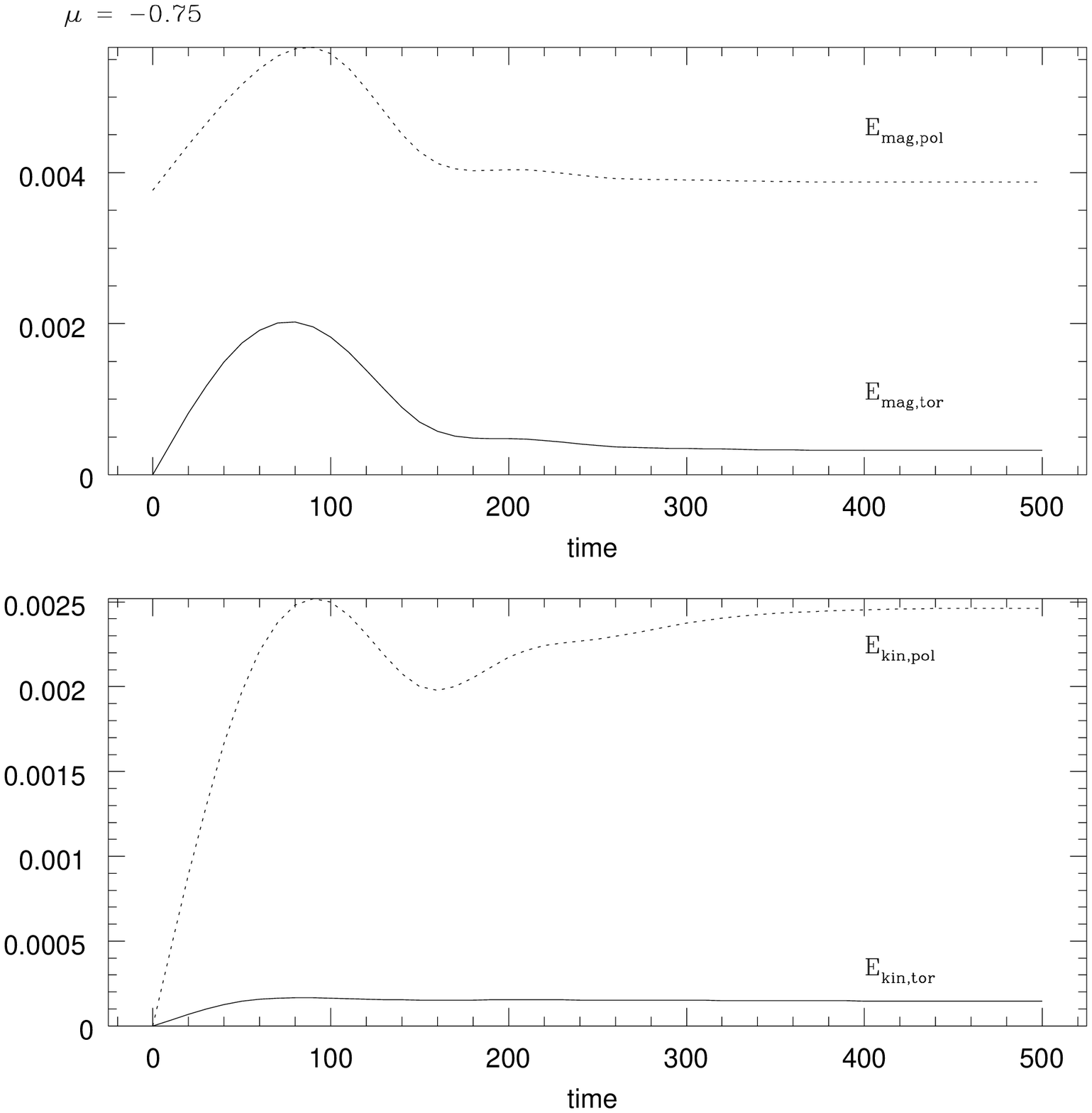}
    }

  \caption{Quantification of the energies involved in the flow.  
	  In the upper/lower panel, the poloidal magnetic/kinetic energy is compared to the 
	toroidal part.  Note that the the poloidal magnetic energy in the last case is double that
	in the other cases.}
  \label{fig7}
\end{figure}

\twocolumn
%\twocolumn

Finally, the ratio of the toroidal 
to poloidal magnetic field 
along our illustrative field line
clearly shows that the predominant magnetic field in 
jets beyond their FM surfaces is the toroidal field
component.  The stability of a jet that is dominated
by a toroidal field is an important problem, and the  
3D simulations needed to investigate
the importance of kink modes are still rare.  Our own 
3D study of a jet in an initially uniform field \citep{Ouyed03}
 shows that jets survive the threatening
non-axisymmetric kink ($m=1$) mode, and we plan to
extend this work  
to these more general configurations. 

\subsection{Jet energetics}

The bulk energetics of our 4 model outflows, 
as a function of time, are shown in 
\ref{fig7a}--\ref{fig7d}.  In every case,
we see that the jets end up being dominated by the poloidal
kinetic
energy of the jet, followed in magnitude by the energy in the 
poloidal magnetic field, and then the energy in the toroidal
magnetic field.  The lowest energy in all of the outflows is
the energy in the overall rotation of the jet.  
Thus, we find 
that 
\begin{equation}
E_{kin,pol} > E_{mag,pol} > E_{mag,tor} > E_{kin,tor}
\end{equation}

These energies shown in this figure are given in 
units of the following physical
quantities: the poloidal and toroidal 
kinetic energies are in units of
$2\pi \rho_i v_{K,i}^2 r_i^3$, while the poloidal and toroidal 
magnetic energies are given in units of $(B_{p,i}^2/2) r_i^3$.

The ratios of the various energy terms are more interesting.
In going through the 4 cases $\mu = 0$ to $ \mu = -3/4$, 
these figures show that the ratio of the poloidal kinetic energy
(bulk energy in outflow) to the bulk energy in the 
poloidal fields, takes the values:

\begin{equation}
{ E_{kin,pol} \over E_{mag,pol}} = 1.5, \quad 2.0, \quad 2.8, \quad 6.3. 
\end{equation}
This shows that the most efficient acceleration of the flows and
the conversion of gravitational binding energy into bulk flow is 
from the less steeply raked mass loads - ie the flows that carry
the most mass.  We noted earlier that the PP92 solutions achieved
the highest terminal Alfv\'en Mach numbers, which goes together
with this result.

A comparison of the energy in the poloidal and toroidal field 
components is most revealing:

\begin{equation}
{E_{mag,pol} \over E_{mag,tor}} = 1.2, \quad 1.3, \quad 2.5, \quad 8.0.  
\end{equation}
Clearly, the relative importance of the
toroidal field becomes progressively less  
as one goes through this mass load sequence - 
exactly as our theoretical model predicts.
The potential and BP82 solutions ($\mu = 0, -0.25$) 
both have toroidal magnetic energies comparable
to their poloidal energies, whereas the other two 
solutions are dominated by the poloidal field component. 
While all four configurations have comparable 
toroidal magnetic energies (with the collimated solutions having more at later
times), the energy of the poloidal magnetic 
field in the last case is almost three times larger than the others.

\subsection{Jet rotation}

\begin{figure*}
        \centering
        \includegraphics[width=1.0\textwidth]{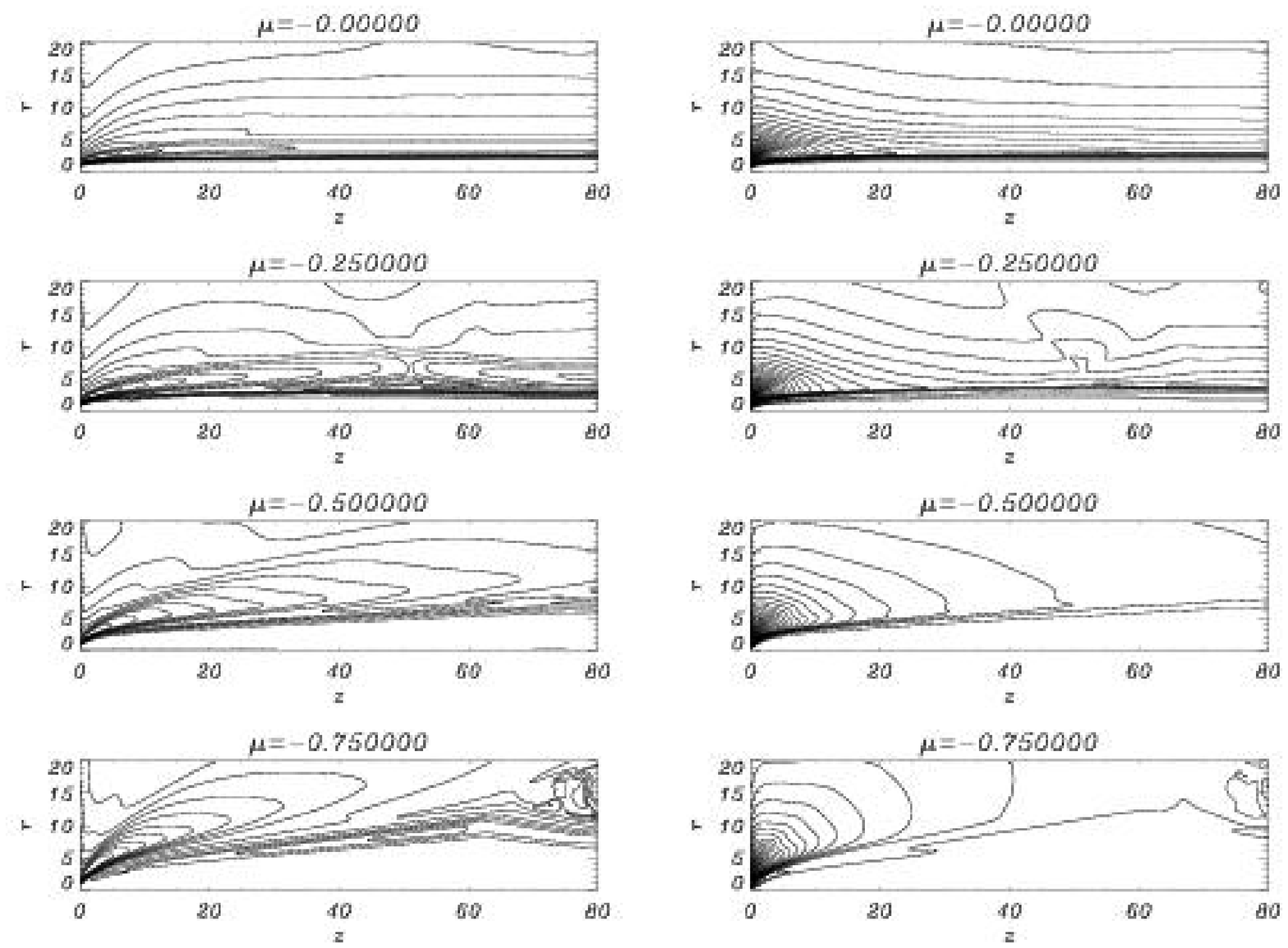}
        \caption[]{Contours of the toroidal velocity (left panels) and toroidal magnetic
	  field (right panels) at $t=400$ for the four cases.  This figure can be compared to
	\ref{fig5a}-\ref{fig5d}.  Note that the magnetic field in the last two panels is too 
	weak to collimate the flow.}
        \label{fig8}
\end{figure*}
%\onecolumn
\begin{figure*}
        \centering
        \includegraphics[width=1.0\textwidth]{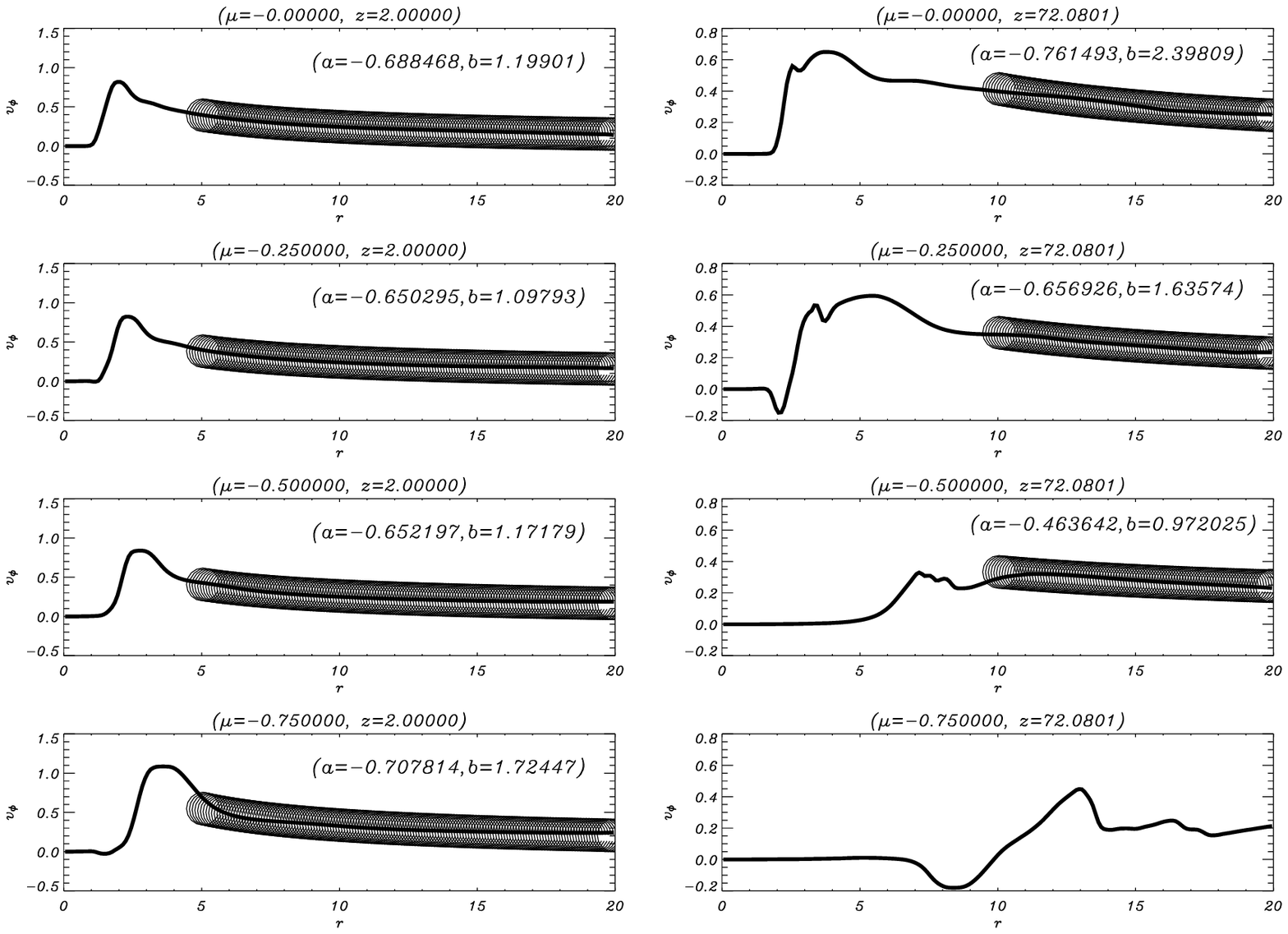}
        \caption[]{Radial profiles of the toroidal velocity within the Alfv\'en surface
	(left panels) and at high $z$ (right panels) for each solution at $t=400$.  The 
	  sub-Alfv\'enic flow can easily be fit by a power law.  At high $z$, only the outermost 
	  radii can be well fit, while the steepest case still exhibits complex behaviour.}
        \label{fig9}
\end{figure*}
%\twocolumn

     The rotational velocity (left hand column) and 
toroidal magnetic field strength (right hand column) 
throughout the 4 outflows, are shown in Figure \ref{fig8}.   
As was found in the potential configurations of OPS97, the top
panel in this figure shows a beautifully, well collimated
outflow with the toroidal field components nearly constant along 
cylinders.  This is also seen for the BP82 case although, 
not as strongly as the potential case.
The wide-angle behaviour of the final two cases is apparent  
in both the toroidal velocity and field contours
as well.  
The explanation for this wide-angle behaviour  
lies in the distribution and strength of the toroidal magnetic field 
(right panels), which is too weak to collimate the flow,
as contrasted with the upper two cases, which clearly has a strong $B_{\phi}$ 
along the jet axis.  

The radial profile of the rotation speed of material in the outflows
are shown in Figure \ref{fig9}.  The two columns show cuts at different
distances $z$ down the jet axis, the right hand column being within
the Alfv\'en surface, while the left hand column is at high z.  At
high z, the outermost radial profiles of the outflow's rotation speed
for the first 3 models can be well fit by power laws;
\begin{equation}
v_{\phi}(r) \propto r^{a}
\end{equation}

\begin{equation}
a=-0.76 \quad -0.66; \quad - 0.46 
\end{equation}
respectively, whereas the 7Q simulation is even 
flatter in profile but too variable to 
fit with a powerlaw at $t=400$. 

These rotational profiles are sufficiently different that 
it may be possible for future observations to place constraints
on the precise nature of the mass load by mapping out the 
radial rotational profile, and using it to infer the underlying
mass load.

\section{Discussion and Conclusions}

We have demonstrated that accretion disks play a major
role in controlling  
the collimation of disk winds through the
radial mass load profiles that they impose on these outflows.  
The conservation laws for stationary, axisymmetric outflows
show that the toroidal field at any point in the jet 
is regulated by this
mass load function.  Given that the toroidal field is responsible for 
the collimating hoop stress in the jet, accretion disks can help 
determine the structure of jets and outflows far from their point
of origin on the disk. 

Our simulations divide into two classes - those that achieve collimation 
towards a cylinder, and those that have a wide-angle structure.  We 
showed that these correspond quite well to predictions 
made by HN89 about jet collimation, 
namely, that if the current $I(r)$ goes to zero in the limit of large
jet radius, that such field lines would have a parabolic structure, whereas
currents that remain finite or diverge result in jets that collimate 
cylindrically.  

We argue that the differences between wide-angle
and highly collimated jets and outflows inferred in the observational
literature, are two aspects of the same underlying mechanism.  Hydromagnetic
disk winds can achieve both types of configuration depending on the 
mass loading that takes place in the underlying accretion disk.
Specifically, we find that the potential and Blandford-Payne 
magnetic field distributions $B_p(r_o) \propto r_o^{-1}, r_o^{-5/4}$
respectively, result in mass loading profiles that produce cylindrically
collimated jets.  The Pelletier-Pudritz configuration on the other
hand, $B_p(r_o) \propto r_o^{-3/2}$ results in a finite current and 
is the boundary between cylindrically and parabolically collimated
outflows.  It may be that nature produces magnetic configurations in 
accretion disks that arise  
through an optimization process (eg. minimum magnetic energy configuration)
of some kind.  We speculate
that it may be that small fluctuations in jet
mass loading, about an energetically preferred 
PP92 profile, would result in jets that are
either cylindrically, or parabollically (ie wide-angle) collimated.
This could explain the variety of jet collimations that are
suggested by the observations. 

Our results allow one to understand why simulation of X-winds by 
 \citet{Shu00}, or split monopoles by Romanova
et al (1997) result in wide-angle flows whose
field lines do not
collimate towards the outflow axis.  
This is a consequence of 
having steep magnetic field gradients, and hence shallow
mass loading profiles which, as we have seen,   
result in weak toroidal fields 
in the jet.  These are unable to collimate the nearly radial
field lines. As \citet{Shu00} note, this does not prevent
the densest material in the outflow from concentrating
in flow that is parallel to the outflow axis.  We see 
this in all of our simulations. 

The radial profiles of various outflow quantities such
as the poloidal outflow speed $v_{pol}$ are tabulated
in this paper.  Our outflows all show the "onion-layer" velocity
structure
that are revealed by the observations of protostellar jets, namely,
that the highest velocity gas is in the interior of the jet
while the lower velocity material are found at increasingly 
larger outflow radii.  Of particular interest is the behaviour of jet
rotation $v_{\phi}(r) \propto r^a$ 
as one goes radially through the jet.
We find that this distribution depends on the underlying mass load
(and hence magnetic configuration in these simulations).  Specifically,
$a= -0.76, -0.66, -0.46$ as one goes from the potential 
configuration, to the Blandford-Payne distribution, and on to the
Pelletier-Pudritz distribution of disk fields $\mu = 0, -1/4, -1/2$
respectively.  

In conclusion, this paper shows that accretion disks can exert a long-range
control of jet properties such as their
collimation and spin, as a consequence of their mass loading. 
What remains to be clarified is exactly
how the accretion disk 
actually does this.  It may
be that the
accretion disk corona plays an important role in this process. 
Dynamical 3D global simulations of the disk and its
outflow, as now being carried out
by an increasing number of authors, 
may offer the best way of exploring the physics of this problem.  

\section*{acknowledgements}

We thank T. Ray and G. Pelletier for stimulating conversations about
these topics, and the organizers and lecturers of 
the Les Houches School on accretion
disks and jets (held in summer 2002) for the stimulating intellectual
atmosphere and incomparable mountain vistas that did much to stimulate
the work in this paper. REP and RO are supported by grants from NSERC
of Canada.
CSR Rogers was supported by McMaster Univeristy and 
wishes to thank 
his parents for their encouragement to pursue his goals and ask questions, 
and $Google^{\texttrademark}$ for answering the questions no one else could.

\appendix

\section{Computing the Initial Configuration of the Magnetic Fields}

   First, we assume axisymmetry, and thus all $\frac{\partial}{\partial \phi}$ terms are ignored.  Second, we are considering only poloidal components; all toroidal magnetic field arises from the centrifugal motion of the disk.  If we take the magnetic field as 
\begin{equation}
\bf{B} = \nabla \times \bf{A}
\end{equation}
then
\begin{equation}
B_r = - \frac{\partial A_{\phi}}{\partial z};
\end{equation}

\begin{equation}
B_{\phi} = 0;
\end{equation}
and 
\begin{equation}
B_z  = -\frac{1}{r} \frac{\partial r A_{\phi}}{\partial r}
\end{equation}

   Thus it remains to find some function $A_{\phi}$ which satisfies our boundary and initial conditions.  We considered only solutions that  were current-free, thus keeping the Lorentz force $ \bf{F} = \bf{J}\times \bf{B} $ on the object also equal to zero.  Since $\bf{J}=\nabla \times \bf{B}$, this means the current-free condition can be written
\begin{equation}
   \frac{\partial B_r}{\partial z} - \frac{\partial B_z}{\partial r} = 0
\end{equation}
which gives us a separable equation for $A_{\phi}$:
\begin{equation}
\frac{\partial^2 A_{\phi}}{\partial r^2} +\frac{1}{r} \frac{\partial A_{\phi}}{\partial r} - \frac{A_{\phi}}{r^2} = - \frac{\partial^2 A_{\phi}}{\partial z^2} = -k^2
\end{equation}
with (non-divergent) basis functions $ A_{\phi, k}$
\begin{equation}
A_{\phi, k} = J_1(kr)\exp(-k|z|)
\end{equation}
yielding a general class of solutions
\begin{equation}
A_{\phi}(r,z) = \int^\infty_0 S(k) J_1(kr) \exp(-k|z|) dk
\end{equation}

   It now remains to find the amplitude $S(k)$, from some boundary condition.  We exploit this freedom and let $B_z$ in the disk have a radial dependence given by:
\begin{equation}
B_z(r,z=0) = b r^{\mu -1}
\end{equation}
where b is a normalizing constant.

Using Eqn A.4 and A.8 and exploiting the properties of Bessel functions, we have the following form for $B_z$:
\begin{equation}
B_z(r,z=0) = \int^\infty_0 S(k) J_0(kr) k dk
\end{equation}
The inverse Hankel Transform gives us an equation for S(k)
\begin{eqnarray}
S(k)  &=& \int^\infty_0 B_z(r,z=0) J_0(kr) r dr \nonumber \\
&=& b \int^\infty_0 r^{\mu} J_0(kr) dr
\end{eqnarray}
which has the solution
\begin{equation}
S(k) = \frac{b}{k^{\mu +1}} f(\mu)
\end{equation}
where
\begin{equation}
f(\mu) = 2^\mu \frac{\Gamma(1/2 + \mu/2)}{\Gamma(1/2 - \mu/2)}
\end{equation}
(Note that this analytic transform was useful for testing our numerical integrator, as explained in the next section.)

Thus, equation A.8 becomes
\begin{equation}
A_{\phi}(r,z) = b f(\mu) \int^\infty_0 \frac{1}{k^{\mu+1}} J_1(kr) \exp(-k|z|) dk
\end{equation}

   This Hankel transform has only two analytical solutions, when $\mu = +1$, which has vertical field lines
 ($B_z = b$, where b is a constant), and when $\mu = 0$, which is the potential solution discussed in OPI.

   We employed a numeric integrator to evaluate three additional configurations, approaching the lower limit of $\mu=-1$; $\mu = -1/4$, which corresponds to the self-similar solutions of BP82, $\mu=-1/2$, the solution suggested by PP92, and $\mu=-3/4$. 

\section{Hankel Transform Algorithm}

Since there are a bevy of Hankel transform algorithms available in 
the public domain, it was unnecessary to devise our own.
  In our search for a numeric integration program capable of 
performing Hankel transforms, we tested several applications. 
 The two most tested programs were designed by \citet{Anderson79,Piessens82}.
The most important criteria for our choice were speed and accuracy.  
The initial conditions for our 500x200 grid were set point by point, 
so it was important that the transforms be done quickly and efficiently. 

Piessens' program 'hankel' is a double-precision code using 
Fast Fourier Transforms (which are done by gaussian
 quadrature methods) to evaluate each transform.  It ran somewhat slowly, 
and the solutions it gave did not behave well near 
the analytic limits ($\mu = -1, +1$) of the functions we were evaluating.  
It was capable of integrating Bessel functions 
of any order, but since our transforms were only of order 0 and 1, this 
feature was unnecessary.
   Anderson's program 'ZHANKS' \citep{Anderson79} is 
a single precision code that uses adaptive filter weights.  The filters are 
designed by using Hankel Transforms with known analytic solutions:

\begin{equation}
\int^\infty_0 \lambda \exp(-a \lambda^2) J_0(b \lambda) d\lambda
= [\exp(-b^2/4a)]/(2a)
\end{equation}
and
\begin{equation}
\int^\infty_0 \lambda^2 \exp(-a \lambda^2) J_1(b \lambda) d\lambda
= b[\exp(-b^2/4a)]/(2a)^2
\end{equation}
where $a > 0, b > 0$.

   Its algorithm uses the fact that algebraically related kernel functions 
(the function being transformed by convolution with 
the Bessel function) transform the same way.  Thus, the computed kernal 
function is saved on its first call, and subsequent 
calculations use the results of previous calculations to evaluate the 
transforms, making computation much faster. 
   The solutions computed by the ZHANKS algorithm are also behaved 
better at the analytic limits, matching closely with the 
solutions we tested using MATLAB.

 In addition to comparisons with known analytic solutions, 
the results of the algorithms were tested in a hydrostatic run of 
our simulation in which the disk did not rotate.  
Since the corona was in pressure balance with the central star and the 
surrounding disk, only magnetic forces would disturb the gas.  
As the initial conditions should be completely 
current-free, no Lorentz forces should be developed, and the 
gas should remain motionless.  This was in fact the case, except 
for slight motions arising from a combination of the accuracy of the 
algorithms and the discrete nature of the grid.  Our 
hydrostatic simulations were run out beyond 400 timesteps for each 
magnetic configuration.  The motions were slightly increased
for the more radially dependent magnetic profiles ($\mu = -1/2$ and $\mu = -3/4)$), 
but in all cases their movement was such that the
hydrostatic balance was disturbed to no more than $5 r_i$ after $400 t_i$

\label{lastpage}
%\bye

\end{document}